\documentclass[aps,prb]{revtex4-1}  
\usepackage[version=3]{mhchem}	
\usepackage{graphicx}  
\usepackage{nicefrac}  
\usepackage{siunitx} 	
\usepackage{amsmath,amssymb,amsfonts}   
\usepackage{hyperref}
\usepackage{import}
\usepackage{mathdesign}
\usepackage{bbold}
\usepackage{wasysym}
\usepackage{xtab}
\usepackage{xtab,afterpage}
\usepackage{lipsum}
\usepackage{multirow}
\usepackage{bm}
\usepackage[normalem]{ulem}

\usepackage{isomath}
\usepackage{bbm}
\usepackage{braket}

\usepackage{xr}
\newcommand{\hmn}[1]{
  \ensuremath{\begingroup\setupHMN #1\endgroup}%
}

\newcommand{\setupHMN}{%
  \doHMN{-}{\HMNoverline}%
  \doHMN{*}{\HMNminverse}%
  \doHMN{`}{\HMNminversep}%
  \doHMN{~}{\HMNminversea}%
  \doHMN{!}{\HMNminversemtilde}%
  \doHMN{i}{\infty}
}

\newcommand{\doHMN}[2]{%
  \begingroup\lccode`~=`#1
  \lowercase{\endgroup\let~}#2%
  \mathcode`#1="8000
}

\newcommand{\HMNminverse}[1]{\frac{#1}{m}}
\newcommand{\HMNminversep}[1]{\frac{#1}{m'}}
\newcommand{\HMNminversea}[1]{\frac{#1}{a}}
\newcommand{\HMNminversemtilde}[1]{\frac{#1}{\tilde{m}}}
\newcommand{\HMNoverline}[1]{\mkern1mu\overline{\mkern-1mu#1\mkern-1mu}\mkern1mu}

\newcommand{\mat}{\matrixsym}
\newcommand{\ten}{\tensorsym}

\newcommand{\mycolor}{black}
\newcommand{\altcolor}{black}

\hypersetup{colorlinks=true,urlcolor=blue,citecolor=blue,linkcolor=blue,breaklinks}
\renewcommand{\vec}[1]{\mathbf{#1}}

\begin{document}


\title{Colour symmetry and altermagnetic-like spin textures in  noncollinear antiferromagnets}

\author{Paolo G. Radaelli}
\affiliation{Clarendon Laboratory, Department of Physics, University of Oxford, Oxford, OX1 3PU, United Kingdom}
\email[Corresponding author: ]{p.g.radaelli@physics.ox.ac.uk}

\author{Gautam Gurung}
\affiliation{Trinity College, University of Oxford, Oxford, OX1 3BH, United Kingdom}
\email[Corresponding author: ]{gautam.gurung@trinity.ox.ac.uk}

\date{\today}

\begin{abstract}

We present a formalism based on colour symmetry to analyse the momentum-space spin textures of non-collinear antiferromagnets.  We show that, out of the spin textures allowed by the magnetic point group, \textcolor{\altcolor} {one can extract a component that is invariant by general rotations in spin space, and can exist in the absence of spin-orbit coupling, in complete analogy to spin textures in altermagnets}.  We demonstrate this approach in the case of three complex, non-collinear magnets, Mn$_3$Ir(Ge,Si), Pb$_2$MnO$_4$ and Mn$_3$GaN.  For Mn$_3$GaN, we also show that the predictions of colour-symmetry analysis are consistent with density functional theory calculations performed on the same system both with and without spin-orbit coupling.

 \end{abstract}

\maketitle

\section{Introduction}

In the past five years, an intense theoretical and experimental effort has been focussed on magnetic systems having electronic bands with lifted spin degeneracy, particularly when this is not due to spin-orbit coupling (SOC) (for example Refs. \onlinecite{Smejkal2022, Smejkal2022b} --- see Ref. \onlinecite{Radaelli2024} for a full list of references).  \textcolor{\altcolor} {Collinear magnets with symmetry-driven compensated net magnetization} displaying this phenomenology have been named `altermagnets', \textcolor{\altcolor} {though this term has been applied by others to a wider group of systems that display spin splitting in momentum space --- see for example Ref. \onlinecite{Cheong2024}}.  \textcolor{\altcolor} {Three} broad classes of magnetic materials are often included in this \textcolor{\altcolor} {extended} classification: altermagnets \textit{stricto sensu}, \textcolor{\altcolor} { non-collinear compensated antiferromagnets which, like altermagnets have spin splitting that is $\vec{k}/-\vec{k}$ \emph{symmetric} and time-reversal odd (TRO)}, and other systems, including so-called $p$-wave magnets,\cite{Hellenes2023} and triangular magnetic lattices \cite{Hayami2020b}, in which spin splitting is $\vec{k}/-\vec{k}$ \emph{anti-symmetric} and time-reversal even (TRE)\textcolor{\altcolor}{\footnote{`$\vec{k}/-\vec{k}$ \emph{symmetric}/\emph{anti-symmetric}' means that the spin texture, defined as $ \vec{s}_{n\vec{k}}=\bra{\Psi_{n\vec{k}}}\bm{\sigma}\ket{\Psi_{n\vec{k}}}$ (see below) is the same/changes sign by exchanging $\vec{k}$ and $-\vec{k}$. `Time-reversal odd/even' means that the spin texture in the time-reversed magnetic domain (i.e., the domain obtained by flipping all the magnetic moments) is opposite/the same for a given $\vec{k}$. }}.  In analogy with the celebrated Rashba-Dresselhaus (R-D) effect,\cite{Dresselhaus1955, Rashba1959} the latter require inversion-symmetry breaking, but unlike R-D systems, do not require SOC.  In a previous paper \cite{Radaelli2024}, one of us (PGR) presented a symmetry classification of $\vec{k}/-\vec{k}$-symmetric, TRO splitting of electronic bands in magnetic materials based on a general tensorial approach \textcolor{\altcolor} {and with a focus on (collinear) altermagnets}.  It was also shown that, for \textcolor{\altcolor} {collinear compensated magnets} where the splitting is allowed \textcolor{\altcolor}{(i.e., those whose symmetries do not contain the time reversal operator or the product of inversion and time reversal operators)}, one can decompose the spin texture into two components: a SOC-independent `altermagnetic' texture, having a general form that can be obtained via the spin-group SG analysis (or an analogue based on black-and-white Shubnikov groups), and a 'residual' component, which is allowed by the exact magnetic point group (MPG) symmetry for a particular direction of the N\'eel vector and will only exist in the presence of SOC.

The existence of non-relativistic spin textures in \emph{non-collinear} antiferromagnets has been discussed before  altermagnetism emerged as a distinct concept, \cite{Zelezny2017}, \textcolor{\altcolor}{while spin-dependent DFT calculation on non-collinear magnets have been performed as far back as  1989. \cite{Sticht1989}.}   Collinear and non-collinear magnets can possess the same MPG symmetries (though cubic MPG can only describe non-collinear magnets), and there is general consensus that the SG analysis, with some modifications, can be extended to non-collinear systems.  Nevertheless, no systematic methods has been proposed to date to extract \textcolor{\altcolor} {SOC-independent} textures in non-collinear systems.  Here, we fill this gap by employing a systematic approach based on so-called colour symmetry groups (CG) to describe $\vec{k}/-\vec{k}$ \emph{symmetric}, time-reversal odd altermagnetic-like textures in collinear and non-collinear magnets.\footnote{As explained in the remainder, the CG approach can be employed to describe the magnetic structures of $\vec{k}/-\vec{k}$-antisymmetric magnets, but coloured tensors as defined here would require some modifications.  For this reason, we will exclude$\vec{k}/-\vec{k}$-antisymmetric magnets from our treatment.}  In Ref. \onlinecite{Radaelli2024}, it was shown that Shubnikov groups offer an alternative formulation of the SG analysis, and offer some advantages --- for instance, they preserve the natural concept of time reversal symmetry.  \textcolor{\altcolor}{The CGs we employ here} are a natural extension of black-and-white Shubnikov groups to cases in which there are more than two spin directions.   Although CGs and SGs are largely equivalent and can both be employed to describe non-collinear SOC-independent textures, CGs are more intuitive, and do not require the rather cumbersome choice of the SG axis orientation.  We will demonstrate the application of this CG analysis to three non-collinear chiral magnets: Mn$_3$Ir(Ge,Si), Pb$_2$MnO$_4$ and  Mn$_3$GaN.  The magnetic structures of Mn$_3$Ir(Ge,Si) and  Pb$_2$MnO$_4$ are described by complex four-colour symmetries, but do not break any crystal symmetries, which means that the CG and MPG analysis are expected to (and do) yield the same textures. Finally, for the well-known non-collinear magnetic antiperovskite Mn$_3$GaN, we present the CG analysis, \textcolor{\altcolor}{including a full tensorial decomposition}, alongside spin-resolved density functional theory (DFT) calculations  with and without SOC.  \textcolor{\altcolor}{In complete analogy with the case of collinear altermagnets,} we show that, in the absence of SOC, the spin texture is exactly as predicted by the CG analysis, while in the presence of SOC, an additional component emerges, which has the more general MPG symmetry and varies depending on the direction of the magnetic moments. 

\section{Altermagnetism and spin-space rotation invariance}
\label{sec: invariance}

The intuitive statement that the SOC-independent component of the spin texture \textcolor{\altcolor}{(the altermagnetic component in collinear systems)} is invariant by rotation in spin space requires some qualification.  In $\vec{k}/-\vec{k}$ \emph{symmetric} TRO altermagnets, the spin texture can be thought as generated by an effective Zeeman field $\vec{B}^{eff} (\vec{k}, n)$, which depends on the wavevector $k$ and on the band index $n$. \cite{Radaelli2024}  \textcolor{\altcolor}{More rigorously, within the context of DFT, for a given wavevector $\vec{k}$ and band index $n$ the spin texture is defined by the vector field  $ \vec{s}_{n\vec{k}}=\bra{\Psi_{n\vec{k}}}\bm{\sigma}\ket{\Psi_{n\vec{k}}}$, where $\bm{\sigma}$ is a vector of Pauli matrices and the integral implied by the $\bra{}$ and $\ket{}$ is over the real-space unit cell.  Although the two fields, though different in magnitude, are equivalent from the symmetry point of view, in this paper we will employ $\vec{s}_{n\vec{k}}$ to make a direct connection with our DFT results (see below). }  Within the tensorial framework, to a given order the spin textures can be expressed as  

\begin{equation} 
\label{eq. tensor_expansion}
\vec{s}_{n\vec{k}} =T^{(l)}_{i,\alpha\beta\gamma \dots} k_\alpha k_\beta k_\gamma \dots
\end{equation}

where $T^{(l)}_{i,\alpha\beta\gamma \dots}$ is tensor of odd rank that is symmetric in the Greek indices. 

If  \textcolor{\altcolor}{$\vec{s}_{n\vec{k}}$} is generated by a collection of \textcolor{\altcolor}{real-space} magnetic moments $\vec{S}_j$ within the crystal structure, \textcolor{\altcolor}{`rotational invariance in spin space' means that, upon a global spin-space rotation of the $\vec{S}_j$'s, the \emph{amplitude}  $|\vec{s}_{n\vec{k}}|$ is invariant, while the vector field direction co-rotates with the $\vec{S}_j$'s.  Mathematically, this can be expressed as}:

\begin{equation}
\label{eq: invariance}
\vec{s}_{n\vec{k}}(\vec{S}_j)=\mat{R}^{-1} \vec{s}_{n\vec{k}}(\mat{R}\,\vec{S}_j)
\end{equation}

where $\mat{R}$ is any rotation matrix.  For collinear magnetic structures, one can verify that the SG construction produces a \textcolor{\altcolor}{$\vec{s}_{n\vec{k}}$} with the invariance properties of eq. \ref{eq: invariance}.  Again, this is most easily seen using the equivalent Shubnikov approach: the \textcolor{\altcolor}{SG-equivalent} Shubnikov group is the symmetry group of the black \& white antiferromagnetic ordering pattern, and is therefore independent on the direction of the N\'eel vector.  As discussed in Ref. \onlinecite{Radaelli2024}, one proceeds to construct a \emph{scalar} field $s_{n\vec{k}}$ from a tensor that is fully symmetrised by the Shubnikov SG, and then equates \textcolor{\altcolor}{$\vec{s}_{n\vec{k}}= \vec{\hat{L}} \, s_{n\vec{k}}$}, where $\vec{\hat{L}} $ is a unit vector in the direction of the N\'eel vector $\vec{L}$.  Since in a collinear structure $\vec{L}$ and the $\vec{S}_j$ are all parallel/antiparallel to each other, eq. \ref{eq: invariance} is obeyed by construction.  Our aim is to find an analogous treatment for $\vec{k}/-\vec{k}$ symmetric, TRO textures in \emph{non-collinear} magnets, so that \textcolor{\altcolor}{$\vec{s}_{n\vec{k}}$} is similarly invariant by construction.  Note that this approach is not immediately suitable for $\vec{k}/-\vec{k}$-antisymmetric magnets, since in this case the spin texture cannot be described as a linear combination of TRO vectors with scalar quantities, which clearly produces a TRO, parity even tensor.  Moreover, in $\vec{k}/-\vec{k}$-antisymmetric magnets, band splitting and spin polarisation are correlated, so that, for a given $k$, the net spin polarisation averaged over bands that would be degenerate in the absence of magnetic ordering is zero.\cite{Brekke2024}

\section{Colour symmetry groups and the description of non-collinear structures}

Colour symmetry groups are groups of combined geometrical and colour-permutational symmetry operations, which leave certain coloured objects unchanged.\cite{Harker1981} In the context of crystallography, one can distinguish between colour \emph{point} groups (CPG), where all geometrical transformations are proper and improper rotations around a fixed point, and colour \emph{space} groups (CSG), in which rotations and translations are combined to yield the symmetries of periodic coloured objects.  Within the CG framework, Shubkikov groups correspond to \emph{bicolour} CGs.  Tricolour CSGs were first classified by D. Harker in 1981,\cite{Harker1981} while other authors later extended the theory to four and six colours.\cite{Sivardiere1984, Roth1985, Sivardiere1988} A full classification of all possible CSGs was presented by J.N. Kozev in 1988.\cite{Kotzev1988}. Applications of CGs to physical problems were discussed as early as 1982, \cite{Litvin1982} and were later extended to include the description of orbital ordering. \cite{Jirak1992} Although extending the Shubkikov group analysis of magnetic structures to multicoloured groups appears quite natural, by the early eighties the representation analysis first introduced by E. F. Bertaut\cite{Bertaut1968} was already very popular, and the alternative CSG approach was not widely pursued.

Here, we will adopt the nomenclature first proposed by Harker to construct multi-colour groups, \cite{Harker1981} in which each CSG/CPG is denoted by an ordinary space or point group, say $G$, followed by two of its subgroups, $H$ and $H'$.  $H'$ is a subgroup of index $n$, where $n$ is also the number of colours, and contains all the operations in $G$ that leave one colour invariant, so that, for example 'red' fragments are transformed in other red fragments.  Other colours are left invariant by conjugated copies of $H'$, which are not necessarily unique to a single colour.  By contrast, $H$ contains all the operations in $G$ that leave \emph{all} colours invariants.  Strictly speaking, adding $H$ to the notation is redundant, because $H$ is the intersection of $H'$ with all its conjugated subsets and can be readily obtained from $H'$ and $G$.  Note also that the colour permutation group is isomorphic to the quotient group $G/H$. Within this framework, the colour symmetry of a particular structure would consist of an ordinary point or space group operation, which only affects the position of the atoms (or fragments), composed with ($\circ$) a colour-permutation operation, which only affects the colours.  So, for example if a certain geometrical operation $g$ converts red (R) fragments to red fragments and exchanges blue (B) and green (G) fragments, the corresponding CG operation is $g \circ \{R\rightarrow R, B\rightarrow G, G\rightarrow B\}$.

When using CGs to describe magnetic structures, each direction of the magnetic moments would be represented by a distinct colour.  It seems useful to consider time reversal as a special operation, (denoted as $1'$), so that structural fragments containing spins related by time reversal are represented by `anti-colours' (e.g., red and anti-red spins would be opposite and related by time reversal).  Using the notation above, $1' \equiv \{R\rightarrow \bar{R}, B\rightarrow \bar{B}, G\rightarrow \bar{G}\}$, where the bar indicates the anti-colour.  An advantage of this approach is that, at any tensor rank, it is possible to represent a colour/anticolour pair with the same coloured tensor, as explained further in Sec. \ref{sec: constructing tensors}.  Moreover, the notion of time reversal is preserved, as in the case of Shubnikov groups.

The exact symmetry group of the crystal + magnetic structure (MPG or magnetic space group --- MSG) consists of all CPG/CSG operations for which the colour permutation corresponds to a geometrical axial vector transformation combined either with the identity or with time reversal.  Hence, the MPG/MSG of a given magnetic structure is a subgroup of its CPG/CSG.  By construction, magnetic structures related by a global rotation in spin space are described by the same CG, the only difference being the identification of colours with spins.

Ever since the work of E. F. Bertaut\cite{Bertaut1968}, irreducible representations of space groups have been the primary tool to describe magnetic structures.  Later, Y. A. Izyumov introduced exchange multiplets, which are sets of irreducible representations that are linked together if the magnetic Hamiltonian is invariant by rotations in spin space.  There is a natural link between CGs and the theory of exchange multiplets \cite{Izyumov1980}, which is discussed in Appendix \ref{sec: Izyumov}.

\textcolor{\mycolor}{\subsection{Colour group notation}
Following Harker,\cite{Harker1981} the CG notation used in the remainder of this paper is $\{G \vert H' \vert H\}$. Although this notation is rather verbose and may be conveniently simplified if the use of colour groups were to become widespread, it has the merit of being complete and accurate and of using only familiar crystallographic symbols.}

\textcolor{\mycolor}{For example, in the case of Mn$_3$GaN (see below), the parent crystallographic group is $G=\hmn{4-3m}$ ; the subgroup $H'$ that leaves one colour invariant is $\hmn{4/mmm}$, and the intersection of the three conjugated  $H'$ groups is the group $H=\hmn{mmm}$.  Hence, the notation for the corresponding 3-color group is $\{\hmn{4-3m} \vert \hmn{4/mmm} \vert\hmn{mmm}\}$.  If the parent group $G$ is centrosymmetric, one can simplify calculations significantly by dropping all improper operators.  In this case, the corresponding colour group is $\{\hmn{432} \vert \hmn{422} \vert\hmn{222}\}$}.

\textcolor{\mycolor}{The notation for the other two systems discussed in this paper (Mn$_3$Ir(Ge,Si) and Pb$_2$MnO$_4$) is obtained in the same way.  In the simplified 4-colour description of Mn$_3$Ir(Ge,Si), each colour is left invariant by a 3-fold axis, so the $H'$ group is $\hmn{3}$, leading to the notation $\{23 \vert 3 \vert 1 \}$ (only the identity operator leaves all colours invariant).  In the 12-colour scheme, no colour is left invariant by operators other than the identity, so the colour group notation is $\{23 \vert 1 \vert 1 \}$.  In the case of Pb$_2$MnO$_4$, the parent group is $\hmn{-42m}$, while the $H'$ subgroup is $m$, with no intersection between the $H'$ groups for different colours other than the identity.  Therefore, the colour group notation is  $\{\hmn{-42m}\vert \hmn{m} \vert \hmn{1}\}$.}

\subsection{Colour groups and spin groups}

Since each crystallographic point group (XPG) is isomorphic to a permutation group, is not surprising that a close relation exist between the CG and SG analyses of both collinear and non-collinear magnetic structures.  This relation becomes transparent if one assigns a distinct colour to each arm of the star generated by the part of the SG that acts exclusively on spins, since the action of the SG on spins becomes equivalent to a colour permutation.  The derivation of all SG  \cite{Litvin1977} is largely identical to that of CPG --- for example, in Ref. \onlinecite{Litvin1977}, the groups $G$ and $H$ are called $R$ and $r$.  However, in the SG derivation, there is an extra step to associate with the quotient group $R/r$ one of the 32 XPGs (called $B$ in  Ref. \onlinecite{Litvin1977}). It should also be noted that CG symbols encode extra information with respect to SGs.  This includes not only the number of colours (i.e., of distinct spin directions), but also the specific representation of the symmetry groups that defines the transformation of all magnetic structures with the same CG symmetry (this is explained in more detail in Appendix \ref{sec: Izyumov}). In the case of SG, the same information is encoded in the orientation of the magnetic moments with respect to the symmetry operators of the group $B$.  This will be further clarified in the examples given below.

\textcolor{\mycolor}{The general group-theoretical statement connecting MPG, SG and CG is that, for a given magnetic structure, the MPG that describes it is a subgroup of the SG, which is itself a subgroup of the CG, though these are not necessarily \emph{proper} subgroups (in other words, pairs of these groups may be isomorphic).   In many cases of interest (including all those discussed in this paper), SGs and the corresponding CGs are also isomorphic, so  whether one uses CG or SG is largely a matter of preference.  However, this is not always the case.  Typically, CG and SG are non-isomorphic when spins connected by colour (permutation) operations do not have the same magnitude, generally as a consequence of a magnetic phase transition that split equivalent sites into inequivalent `orbits'. An example of this is discussed in the Supplementary Information (SI), section I.  When the CG is a proper supergroup of the SG, the CG analysis is more general, and may unveil `hidden' symmetries that are invisible to the SG analysis.  }

 We also find the CG approach to be somewhat more intuitive: \textcolor{\altcolor}{in the case of SG and} for a generic orientation of the magnetic moments (e.g., in the presence of an applied magnetic field), the point group elements acting on spins are generally not co-aligned with those acting on atoms, and their operators do not coincide with any of the symmetry operators that are present in the paramagnetic structure.  By contrast, spin (or colour) permutation has the same meaning regardless of the overall spin orientation.  An example of parallel SG and CPG analyses will be given in Sec. \ref{sec: SG CPG parallel}.

\section{Constructing coloured altermagnetic tensors}
\label{sec: constructing tensors}

The procedure to construct altermagnetic tensors in the framework of CG theory is rather straightforward, and follows closely the case of collinear structures (bicolour groups).  Using the projection method, one constructs sets of \emph{coloured tensors} that are totally symmetric by the CPG of the magnetic structure.  This is done beginning with the most generic form of a symmetric tensor of a given rank, and assigning to it a single colour (say, red).  The `red' tensor is then constructed by adding up all copies of the original tensor transformed in the usual way via symmetry operations in $H'$, where $H'$ transforms red fragments into other red fragments, and normalising by the order of $G$.  Likewise, the `blue' tensor is obtained by adding up all copies of the original tensor transformed via $g\circ H'$, where $g$ transforms red fragments into blue fragments, etc.  The full altermagnetic tensor is then obtained as a linear combination of the coloured tensors \emph{times} the axial unit vector corresponding to the colour assigned to the magnetic moment in the real-space structure.  Once again, by construction, effective magnetic fields (i.e., spin textures) calculated using this method are invariant by eq. \ref{eq: invariance}, since magnetic moments in real space and axial unit vectors in reciprocal space are parallel and hence co-rotating.   Moreover, since the MPG of the magnetic structure is a \emph{subgroup} of the CPG, it follows that the altermagnetic tensor thus constructed is always a \emph{special case} of the general MPG tensor, which can then be decomposed into `altermagnetic-like' and residual components, precisely as in the collinear case.

The procedure to obtain coloured tensors is slightly modified in the presence of geometrical operators composed with the time reversal operator $1'$, which gives rise to anti-colours.  One can obtain separate tensors for a given colour (say, red) and its anti-colour (anti-red).  However, when the full tensor is reconstructed, the red and anti-red tensors will be multiplied by antiparallel vectors, which is equivalent to multiplying the `red' vector by the \emph{difference} of the two tensors.  Using this method, one effectively halves the number of required tensors.  This mirrors the procedure outlined in Ref. \onlinecite{Radaelli2024} for collinear structures, where only one TRO tensor was required rather than separate `black' and `white' tensors.  This situation is discussed further in Sec. \ref{sec: Pb2MnO4}

\section{M\lowercase{n}$_3$I\lowercase{r}(G\lowercase{e},S\lowercase{i})}
\label{sec: 4_colours}

Mn$_3$Ir(Ge,Si) crystallises in the crystallographic space group $\hmn{P2_13}$.  Below the N\'eel temperature (T$_N$=225 K for Ge and 210 K for Si), magnetic moments on the 12 magnetic sites order forming a complex non-collinear structure with the MSG also being $\hmn{P2_13}$ (MPG $23$). \cite{Eriksson2004a, Eriksson2004}  Spin textures and bulk properties in Mn$_3$IrSi  were recently analysed theoretically and discussed in the context of the Landau theory. \cite{Hu2024}

\begin{figure}[!h]
\centering
\includegraphics[scale=0.5]{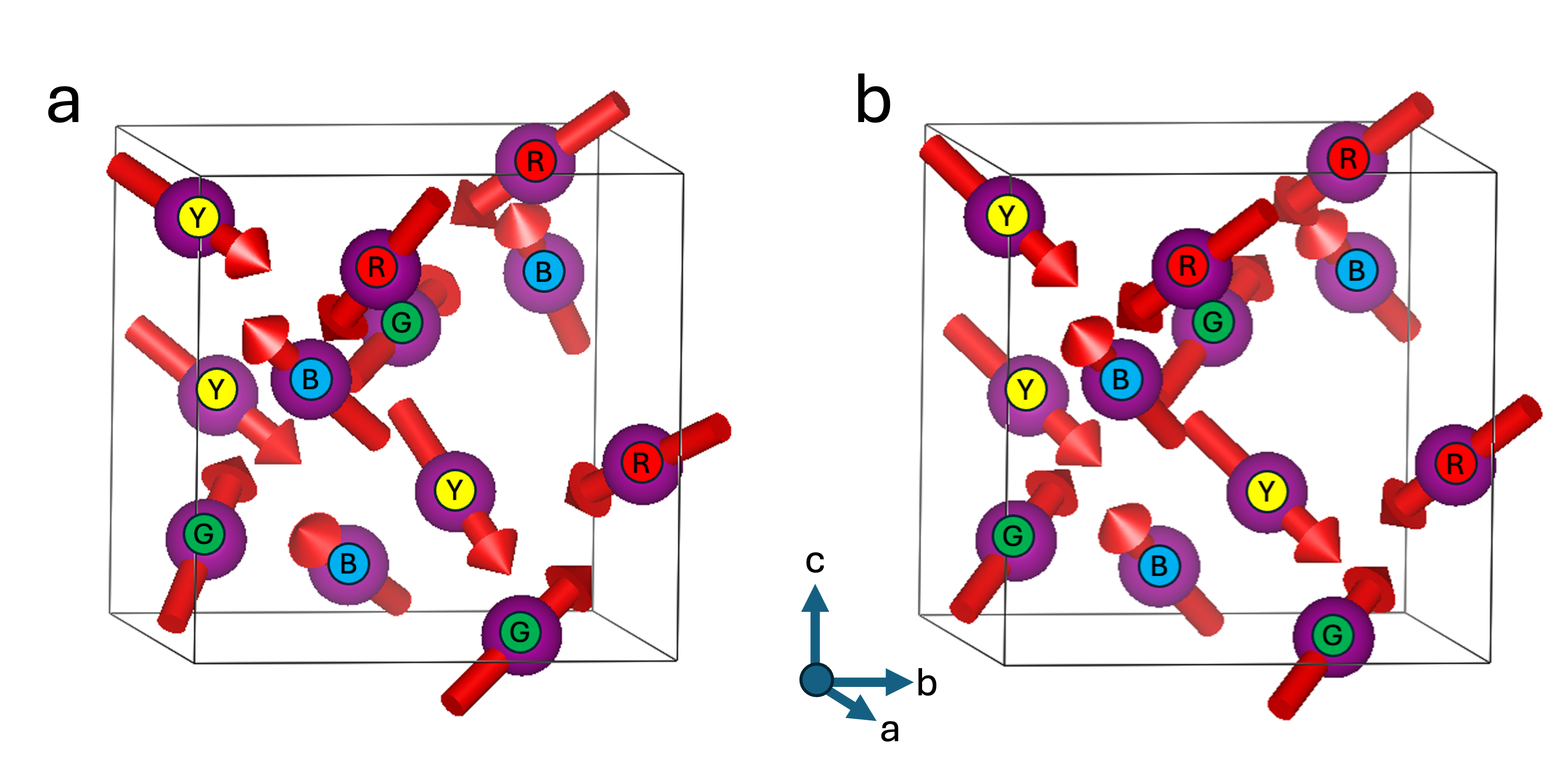}
\caption{ \label{fig: Mn3IrGe_structures}  (Colour online) Experimentally determined \cite{Eriksson2004a} (a) and approximate (b) magnetic structures of Mn$_3$Ir(Ge,Si).  Only magnetic Mn atoms are shown.  Labels indicate distinct colours (see text for the assigment of magnetic moment directions to colours).  }
\end{figure}

In the magnetic structure refined from experimental data and also confirmed by theoretical calculations, \cite{Eriksson2004a, Eriksson2004} all the magnetic moments on the 12 sites are non-collinear, \textcolor{\altcolor}{so analysing the magnetic structure with CG-SG would require 12 colours or 12 distinct spin-space rotations}.  However, groups of three moments are roughly parallel to each other and to one of the cube diagonals, the angle with the closest diagonal being $\sim 11^\circ$.  The experimental magnetic structure (a) and the approximate structure (b) are displayed in Fig. \ref{fig: Mn3IrGe_structures}.  We begin by performing an analysis of the spin textures for the approximate structure, which requires four colours.  Later, we will explain that the 12-colour analysis yields an identical result.  Since the magnetic and crystallographic space groups are the same, the magnetic structure does not break any crystallographic symmetry.  We therefore expect the textures arising from the 4- and 12-colour analysis to be fully consistent with the MPG textures, which are reported in Table I of Ref. \onlinecite{Radaelli2024} (MPG $\hmn{23}$, Class XVII).

\subsection{M\lowercase{n}$_3$I\lowercase{r}(G\lowercase{e},S\lowercase{i}): four-colour analysis}

We choose the four colours to correspond to the following spin directions (see Fig. \ref{fig: Mn3IrGe_structures}(b)):

\begin{eqnarray}
\label{eq: 4_colours}
\rm{Yellow (Y):}&\rightarrow&(1,1,-1)\nonumber\\
\rm{Blue (B):}&\rightarrow&(1,-1,1)\nonumber\\
\rm{Green (G):}&\rightarrow&(-1,1,1)\nonumber\\
\rm{Red (R):}&\rightarrow&(-1,-1,-1)
\end{eqnarray}

The XPG ($\hmn{23}$) comprises 12 symmetry operations: 8 3-fold ($\pm$120$^\circ$) rotations, around $[1,1,1]$ and equivalent directions, three 2-fold rotations around $[1,0,0]$ and equivalent directions and the identity.  The MPG operations that leave Y invariant are the identity and the $\pm120^\circ$ rotations around $[1,1,-1]$, corresponding to point group $\hmn{3}$.  Likewise B,G, and R are left invariant by a conjugated point group, also with symbol $\hmn{3}$ but with a different rotation axis.  There are four such conjugated subgroups in the MPG $\hmn{23}$, and the only common operation to all is the identity.  Hence, the full CPG symbol is $\{\hmn{23} \vert \hmn{3} \vert \hmn{1}\}$.  In this  CPG, 3-fold rotations leave one colour invariant and permute the other three, while two-fold rotations exchange pairs of colours --- for example, the two-fold rotation around $[1,0,0]$ exchange Y with B and G with R.

The next step is to construct four `coloured scalar' tensors of the selected rank using the projection method.  This is done starting from the most generic form of the symmetric scalar tensor, to which we assign the colour Y.  We then apply the space transformation operations of the XPG and exchange the colours as explained in Sec. \ref{sec: constructing tensors}.  We perform this operation for the lowest rank (rank 2), but the procedure is identical for any rank.  The most generic rank-2 symmetric tensor is:

\begin{equation}
\label{eq: ten_gen}
\ten{T}_g=\left(
\begin{array}{ccc}
a & d & e  \\
 d & b & f  \\
 e & f & c  \\
\end{array}
\right)
\end{equation}

where the scalar quadratic form  is obtained by multiplying this matrix to the left and right by $[k_x, k_y, k_z]$.  The four coloured tensors obtained by the projection methods are:

\begin{eqnarray}
\label{eq: ten_colour}
\ten{T}_Y&=\frac{1}{12}\left(
\begin{array}{ccc}
A & B & -B  \\
 B & A & -B  \\
 -B & -B & A  \\
\end{array}
\right)\nonumber\\
\ten{T}_B&=\frac{1}{12}\left(
\begin{array}{ccc}
A & -B & B  \\
 -B & A & -B   \\
 B & -B  & A  \\
\end{array}
\right)\nonumber\\
\ten{T}_G&=\frac{1}{12}\left(
\begin{array}{ccc}
A & -B & -B  \\
 -B & A & B  \\
 -B & B & A  \\
\end{array}
\right)\nonumber\\
\ten{T}_R&=\frac{1}{12}\left(
\begin{array}{ccc}
A & B & B  \\
 B & A & B  \\
 B & B & A  \\
\end{array}
\right)\nonumber\\
\end{eqnarray}

with $A=a+b+c$, $B=1/2(d-e-f)$.  With this construction, the set $\{\ten{T}_Y, \ten{T}_B, \ten{T}_G, \ten{T}_R\}$ is totally symmetric by the CPG --- in other words, it is invariant by all geometrical transformations composed with the colour permutation operations.

To obtain the spin texture at the lowest rank, we re-assign colours to spin texture directions (Eq. \ref{eq: 4_colours}) and add up all the coloured tensors, obtaining the full rank-3 CPG tensor

\begin{equation}
\label{eq: tensor_recomp_4c}
\ten{T}_{\rm{CPG}}=(1,1,-1) \ten{T}_{Y}+(1,-1,1)\ten{T}_{B}+(-1,1,1) \ten{T}_{G}+(-1,-1,-1) \ten{T}_{R}
\end{equation}

It is apparent that the texture obtained from Eq. \ref{eq: tensor_recomp_4c} is completely invariant by rotation in spin space in the sense of Eq. \ref{eq: invariance}.  The expression of $\ten{T}_{\rm{CPG}}$ as a $3\times 6$ matrix is

\begin{equation}
\label{eq: tensor_recomp_4c_3x6}
\ten{T}_{\rm{CPG}}=\left(
\begin{array}{cccccc}
 0 & 0 & 0 & \Lambda _{14} & 0 & 0 \\
 0 & 0 & 0 & 0 & \Lambda _{14} & 0 \\
 0 & 0 & 0 & 0 & 0 & \Lambda _{14} \\
\end{array}
\right)
\end{equation}

with $\Lambda _{14} =1/3(-d+e+f)$ and is formally identical to the one obtained from the MPG analysis, which is reported in Table I of Ref. \onlinecite{Radaelli2024} (MPG $\hmn{23}$, Class XVII), exactly as we expected.

\begin{figure}[!h]
\centering
\includegraphics[scale=0.5]{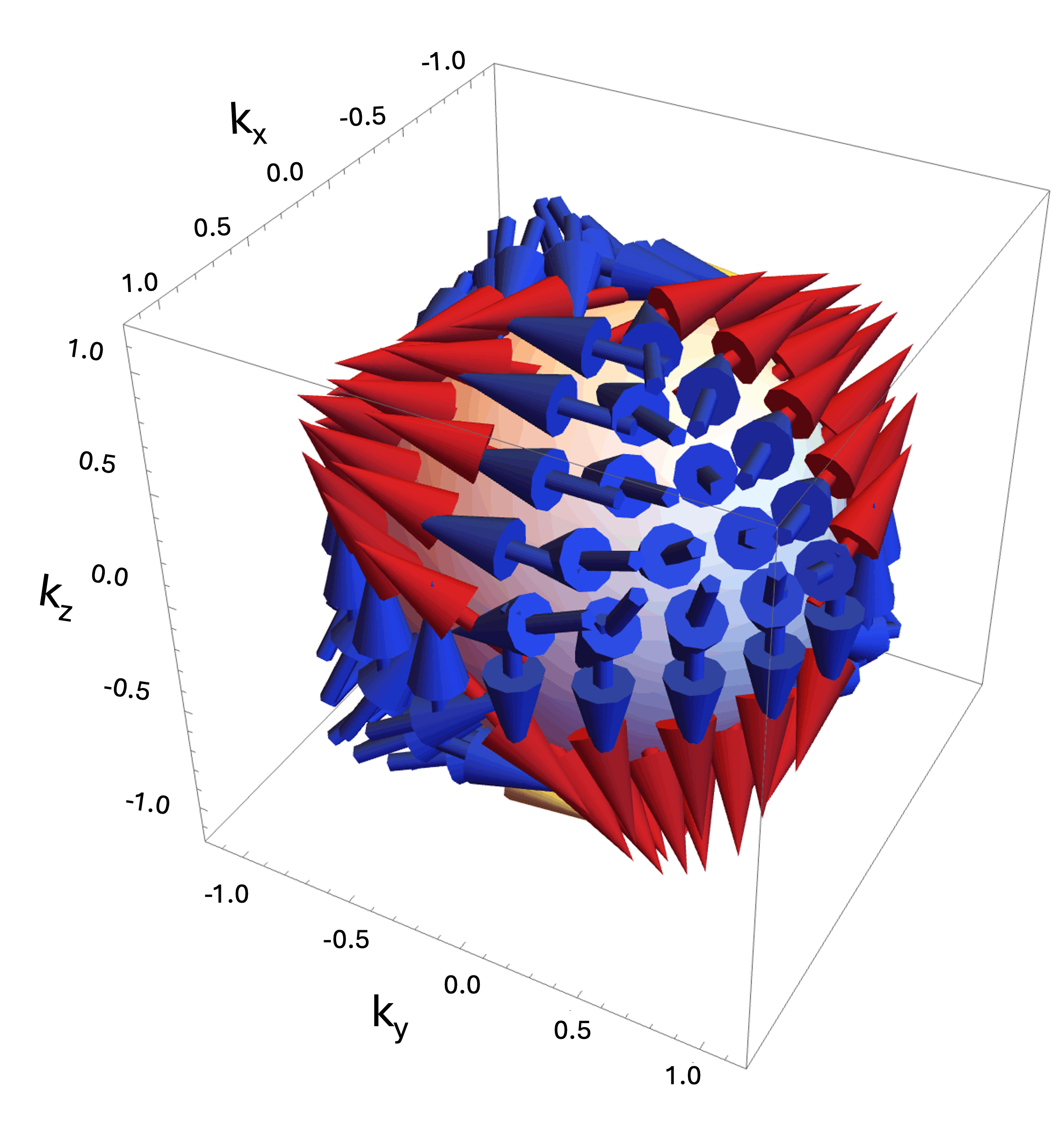}
\caption{ \label{fig: Mn3IrGe_texture}  (Colour online) Spin texture of Mn$_3$Ir(Ge,Si), generated from the tensor in Eq. \ref{eq: tensor_recomp_4c_3x6}.  The colour of the arrows indicates the radial projection of the texture (red=out; blue=in). Wavevectors are in arbitrary units.}
\end{figure}

\subsection{M\lowercase{n}$_3$I\lowercase{r}(G\lowercase{e},S\lowercase{i}): 12-colour analysis}
\label{sec: 12_colours}

We can repeat the same analysis using the experimental magnetic structure of Mn$_3$Ir(Ge,Si), which requires twelve colours, none of which is left invariant by operations other than the identity, hence the 12-colour CPG is $\{\hmn{23} \vert \hmn{1} \vert \hmn{1}\}$.  Here, each of the four colours in the simplified magnetic structure is split into three, so that, for example:

\begin{eqnarray}
\label{eq: 12_colours}
\rm{Yellow 1 (Y1):}&\rightarrow&(v_x,v_y,v_z)\nonumber\\
\rm{Yellow 2 (Y2):}&\rightarrow&(v_y,-v_z,-v_x)\nonumber\\
\rm{Yellow 3 (Y3):}&\rightarrow&(-v_z,v_x,-v_y)\nonumber\\
\end{eqnarray}

etc., where, in the experimental structure, $v_x=0.7106, v_y=1.2118, v_z=-1.015$.  The analysis proceeds in the same way as for the 4-colour structure, yielding the same result (Eq. \ref{eq: tensor_recomp_4c_3x6}), this time with  $\Lambda _{14} =1/3(v_x f+v_y e+v_z d)$.  Note the two results coincide for $\vec{v}=(1,1,-1)$.

In conclusion, for Mn$_3$Ir(Ge,Si), both the 4- and the 12-colour analysis produce tensorial spin textures that are fully consistent with the MPG analysis (Ref. \onlinecite{Radaelli2024}), as we expected since the magnetic structure does not break any of the crystallographic symmetry operators.  We generally expect this to be the case at every tensor order, whenever the `grey' (paramagnetic) group corresponding to the MPG is identical to the XPG.  The CPG tensorial spin texture is formally invariant by a global rotation of the magnetic structure in spin space, and this remains true for spin rotations that break one or more of the MPG $\hmn{23}$ symmetries.  If such low-symmetry magnetic structure could be stabilised, we would expect the altermagnetic-like component of the spin texture (given by the CPG analysis) to remain the same, while an additional component would be activated in the presence of SOC, with tensor forms consistent with the new, lower-symmetry MPG.  However, to our knowledge, no such structure has been observed experimentally.

\section{Dealing with anti-colours: the case of  \textsc{P\lowercase{b}$_2$M\lowercase{n}O$_4$}}
\label{sec: Pb2MnO4}

Pb$_2$MnO$_4$ received some attention due to its potential multiferroic properties. \cite{Kimber2007, Kakarla2019}, \textcolor{\altcolor}{and was recently discussed in Ref. \onlinecite{Cheong2025} as having properties that are consistent with momentum-space spin splitting.}  It crystallises with a  tetragonal structure (space group: \hmn{P-42_1c}) and orders non-collinearly at the $\Gamma$ point below $\sim$18 K, with the magnetic moments on the eight Mn sites aligned along the diagonals of the square faces (Fig. \ref{fig: Pb2MnO4_fig}).  The MSG and MPG are \hmn{P-4'2_1c'} and \hmn{-4'2m'}, respectively.  As in the case of Mn$_3$Ir(Ge, Si), magnetic ordering does not break any crystal symmetry operator, so we expect the CPG and MPG analyses (the latter reported in in Table I, Ref. \onlinecite{Radaelli2024} as Class X) to yield identical results.  However, the tensor associated with Class X has two parameters ($\Lambda_{14}$ and $\Lambda_{36}$), so the specific values of these parameters in the CPG analysis will depend on the relative orientations of the spins in real space, whilst remaining invariant by global rotations in spin space.

The colours/anticolours have been associated with magnetic moment directions, as follows:

\begin{figure}[!h]
\centering
\includegraphics[scale=0.5]{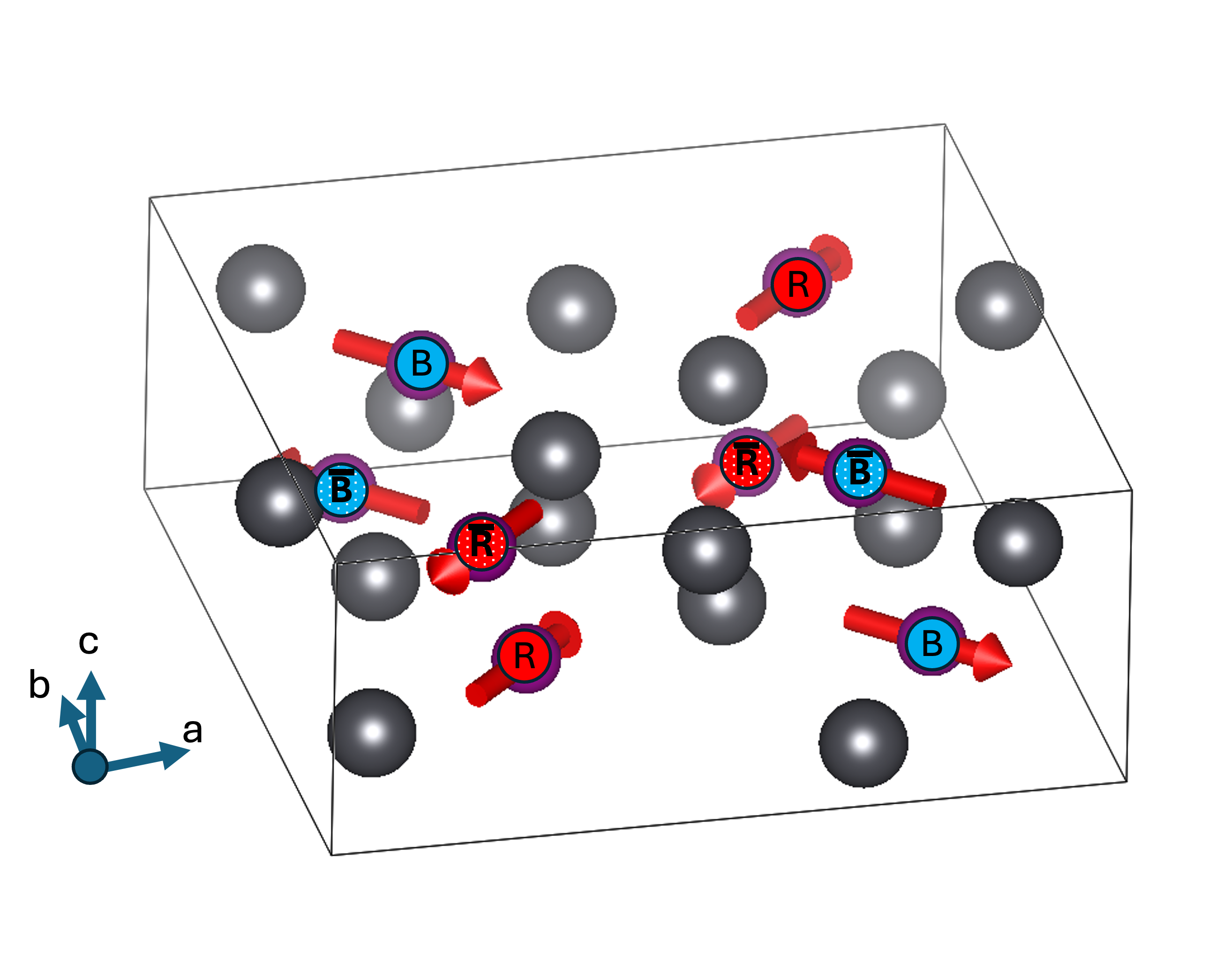}
\caption{ \label{fig: Pb2MnO4_fig}  (colour online) Magnetic structure of Pb$_2$MnO$_4$, based on Ref. \onlinecite{Kimber2007}.  The magnetic Mn atoms have associated arrows, while the grey spheres are Pb (oxygen atoms have been omitted).  Colours and anti-colours are indicated with plain and dotted fields, respectively. } 
\end{figure}

\begin{eqnarray}
\label{eq: anticolours}
\rm{Red (R):}&\rightarrow&(1,1,0)\nonumber\\
\rm{Blue (B):}&\rightarrow&(1,-1,0)\nonumber\\
\rm{anti-Red (\bar{R}):}&\rightarrow&(-1,-1,0)\nonumber\\
\rm{anti-Blue (\bar{B}):}&\rightarrow&(-1,1,0)\nonumber\\
\end{eqnarray}

There are eight symmetry operators in the XPG --- the identity ($E$), two $\bar{4}$ operations, a 2-fold rotation around $z$, two 2-fold axes along $x$ and $y$ and two mirror planes perpendicular to $xy$ and $\bar{x}y$.  The corresponding colour permutation operations are the following:

\begin{eqnarray}
\label{eq: anticolour - ops}
E&:& \;\;\;R\rightarrow R; B\rightarrow B;\bar{R}\rightarrow \bar{R};\bar{B}\rightarrow \bar{B} \nonumber\\
\bar{4}^+&:& \;\;\;R\rightarrow B; B\rightarrow \bar{R};\bar{R}\rightarrow \bar{B};\bar{B}\rightarrow R \nonumber\\
\bar{4}^-&:& \;\;\;R\rightarrow \bar{B}; B\rightarrow R;\bar{R}\rightarrow B;\bar{B}\rightarrow \bar{R} \nonumber\\
2_z&:& \;\;\;R\rightarrow \bar{R}; B\rightarrow \bar{B};\bar{R}\rightarrow R;\bar{B}\rightarrow B \nonumber\\
2_x&:& \;\;\;R\rightarrow \bar{B}; B\rightarrow \bar{R};\bar{R}\rightarrow B;\bar{B}\rightarrow R \nonumber\\
2_y&:& \;\;\;R\rightarrow B; B\rightarrow R;\bar{R}\rightarrow \bar{B};\bar{B}\rightarrow \bar{R} \nonumber\\
m_{xy}&:& \;\;\;R\rightarrow \bar{R}; B\rightarrow B;\bar{R}\rightarrow R;\bar{B}\rightarrow \bar{B} \nonumber\\
m_{\bar{x}y}&:& \;\;\;R\rightarrow R; B\rightarrow \bar{B};\bar{R}\rightarrow \bar{R};\bar{B}\rightarrow B \nonumber\\
\end{eqnarray}

Consequently, the $H'$ subgroup (for red and anti-red) is $m=\{E,m_{\bar{x}y}\}$, while the group $H$ only contains the identity.  The CPG symbol is $\{\hmn{-42m}\vert \hmn{m} \vert \hmn{1}\}$.  The subgroup $m$ has index 4 in \hmn{-42m}, so $\{\hmn{-42m}\vert \hmn{m} \vert \hmn{1}\}$ is a four-colour group.

The four coloured tensors obtained by the projection methods are:

\begin{eqnarray}
\label{eq: ten_col_acol}
\ten{T}_R&=\frac{1}{8}\left(
\begin{array}{ccc}
A & D & B  \\
D & A & B  \\
 B & B & C  \\
\end{array}
\right)\nonumber\\
\ten{T}_{\bar{R}}&=\frac{1}{8}\left(
\begin{array}{ccc}
A & D & -B  \\
D & A & -B  \\
 -B & -B & C  \\
\end{array}
\right)\nonumber\\
\ten{T}_B&=\frac{1}{8}\left(
\begin{array}{ccc}
A & -D & B  \\
-D & A & -B  \\
 B & -B & C  \\
\end{array}
\right)\nonumber\\
\ten{T}_{\bar{B}}&=\frac{1}{8}\left(
\begin{array}{ccc}
A & -D & -B  \\
-D & A & B  \\
 -B & B & C  \\
\end{array}
\right)
\end{eqnarray}

with $A=a+b$, $B= (ee + f)/2$, $C=2c$, and $D=d$.  We can combine each colour with its anti-colour as:

\begin{eqnarray}
\label{eq: ten_col_sub_acol}
\ten{T}_{R-\bar{R}}&=\frac{1}{4}\left(
\begin{array}{ccc}
0 & 0 &  B  \\
0 & 0 & B  \\
 B & B & 0  \\
\end{array}
\right)\nonumber\\
\ten{T}_{B-\bar{B}}&=\frac{1}{4}\left(
\begin{array}{ccc}
0 & 0 & B  \\
0 & 0 & -B  \\
 B & -B & 0  \\
\end{array}
\right)
\end{eqnarray}

Finally, we can reconstruct the full tensor using the colour assignments in eq. \ref{eq: anticolours}: 

\begin{equation}
\label{eq: tensor_recomp_anticol}
\ten{T}_{\rm{CPG}}=\left(
\begin{array}{cccccc}
 0 & 0 & 0 & \Lambda _{14} & 0 & 0 \\
 0 & 0 & 0 & 0 & \Lambda _{14} & 0 \\
 0 & 0 & 0 & 0 & 0 & 0 \\
\end{array}
\right)
\end{equation}

where  $\Lambda _{14}=B$.  This is to be compared with the MPG tensor for Class X:

\begin{equation}
\label{eq: tensor_recomp_anticol_MPG}
\ten{T}_{\rm{MPG}}=\left(
\begin{array}{cccccc}
 0 & 0 & 0 & \Lambda _{14} & 0 & 0 \\
 0 & 0 & 0 & 0 & \Lambda _{14} & 0 \\
 0 & 0 & 0 & 0 & 0 & \Lambda _{36}  \\
\end{array}
\right)
\end{equation}

so $\ten{T}_{\rm{CPG}}$ is a particular case of $\ten{T}_{\rm{MPG}}$, while $\Lambda_{36}=0$ clearly arises from the fact that all magnetic moments are in the $xy$ plane.  One can easily verify that $\ten{T}_{\rm{CPG}}$ is invariant by rotation in spin space in the sense of Eq. \ref{eq: invariance}.   It can also be shown that more general four-colour models based on the same CPG produce an altermagnetic-like tensor that is identical to $\ten{T}_{\rm{MPG}}$.

\section{Breaking crystal symmetry with magnetic moments: the case of \textsc{M\lowercase{n}$_3$G\lowercase{a}N}}

\begin{figure}[!h]
\centering
\includegraphics[scale=0.5]{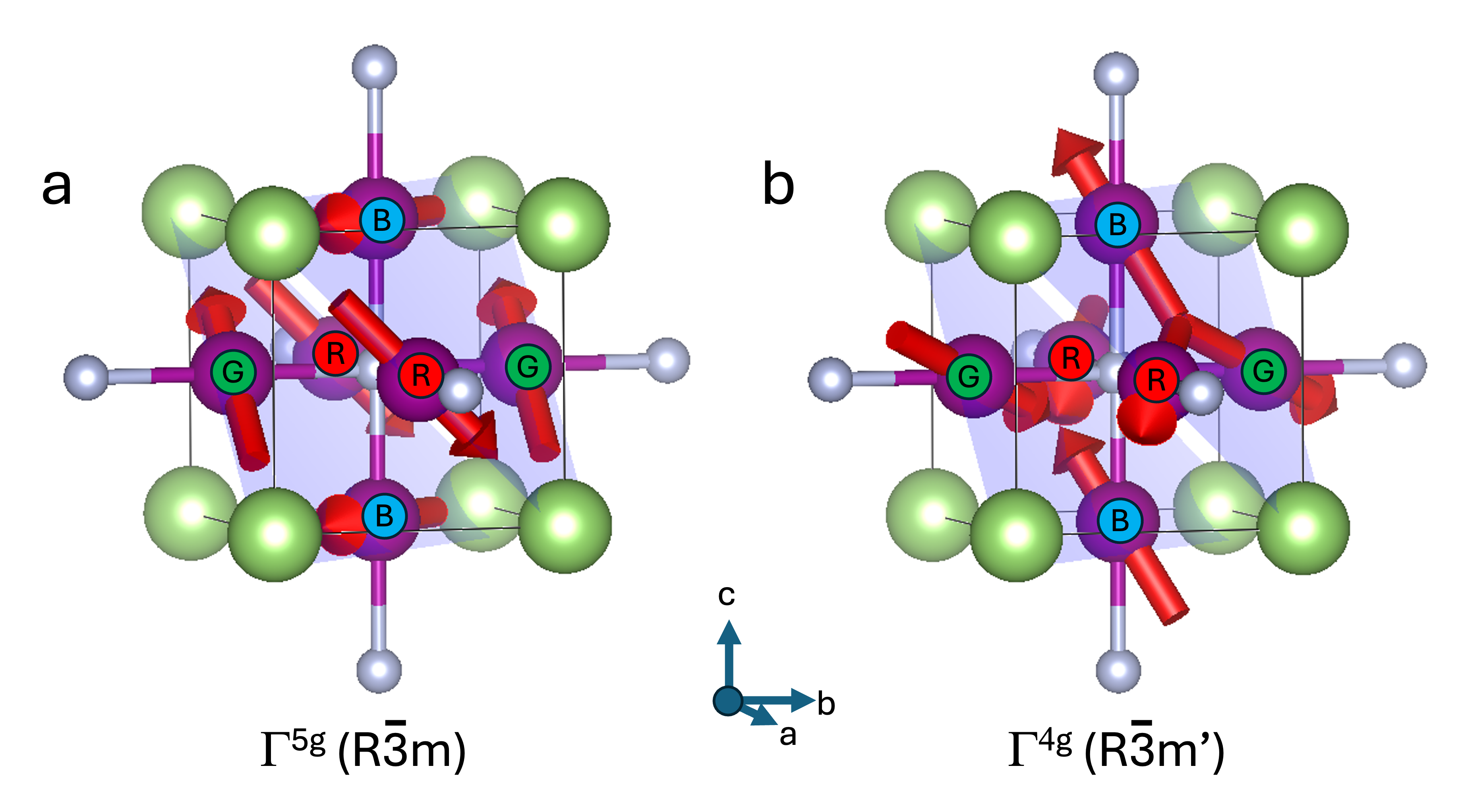}
\caption{ \label{fig: Mn3GaN_structures}  Magnetic structures of the $\Gamma^{5g}$ (a) and $\Gamma^{4g}$ (b) phases of Mn$_3$GaN. \cite{Bertaut1968b} Magnetic Mn ions are shown with arrows, small spheres are N and large spheres are Ga.  (111) planes containing the Mn kagom\'e lattice are shadowed.}
\end{figure}

The cases of Mn$_3$Ir(Ge,Si) and Pb$_2$MnO$_4$ were complex and instructive, but mainly served as validations of the colour symmetry approach, since there was an expectation that the results should coincide with the MPG analysis.  Cases in which the crystal symmetry is explicitly broken by the magnetic structure are significantly more interesting, particularly when the phase diagram contains phases with different magnetic symmetries.  The  anti-perovskite family, of which Mn$_3$GaN is a well-known representative,\cite{Bertaut1968b, Shi2016} is an ideal case in point.   In the bulk, Mn$_3$GaN orders with a non-collinear AFM structure and T$_N \sim$ 290 K, but in thin films the N\'eel temperature can be as high as T$_N \sim$ 350 K, making this material suitable for applications in spintronics.  Since the magnetic Mn sites form kagom\'e lattices with Ga at the empty sites, the magnetic ordering is in a typical 120$^\circ$ pattern, with the magnetic and crystallographic unit cells coinciding ($\Gamma$-point ordering). The crystallographic space group is $\hmn{Pm-3m}$ (No. 221).  Two magnetic structures have been proposed for Mn$_3$GaN, corresponding to the $\Gamma^{5g}$ and $\Gamma^{4g}$ \emph{irreps} of $\hmn{Pm-3m}$, and having MSGs $\hmn{R-3m}$ and $\hmn{R-3m'}$, respectively (see Fig. \ref{fig: Mn3GaN_structures}).  Note that in both cases the cubic crystal symmetry is broken by the magnetic structure.  Mn$_3$GaN itself appears to be ordering with $\Gamma^{5g}$ in both bulk\cite{Bertaut1968b} and thin film\cite{Nan2020} forms, but DFT calculations\cite{Singh2021} have shown that $\Gamma^{4g}$ can be stable for other members of the antiperovskite family,\textcolor{\altcolor}{\footnote{Note that the $\Gamma^{4g}$ magnetic structure is the same as for the compound Mn$_3$Ir reported in ref. \onlinecite{Zelezny2017}.  The schematic spin texture reported in their Fig. 3 is very similar to the one we calculate here.}} sometimes associated with a $\Gamma^{5g}/\Gamma^{4g}$ phase transision.   As drawn in Fig. \ref{fig: Mn3GaN_structures}, the magnetic moments in the two structures are related by a 90$^{\circ}$ collective rotation of the magnetic moments in spin space.  However, the MPG $\hmn{-3m'}$ of the $\Gamma^{4g}$ is \emph{admissible}, i.e., it allows the development of SOC-induced weak ferromagnetism via collective tilting of the moments towards the $[1,1,1]$ direction.

\subsection{\textsc{M\lowercase{n}$_3$G\lowercase{a}N}: CPG analysis}

The XPG $\hmn{m-3m}$ comprises 48 symmetry operations, but improper operations have the same effect on magnetic moments as the corresponding proper rotations, so it will be sufficient to consider the 24 operations of point group $\hmn{432}$:  eight 3-fold ($\pm$120$^\circ$) rotations, around $[1,1,1]$ and equivalent directions, three 2-fold rotations and six 4-fold rotations around $[1,0,0]$ and equivalent, six 2-fold rotations around $[1,1,0]$ equivalent, and the identity.  The colour assigment (shown in Fig. \ref{fig: Mn3GaN_structures}) is the same for the $\Gamma^{5g}$ and $\Gamma^{4g}$ structures, except for a collective rotation in spin space.  For $\Gamma^{5g}$:

\begin{eqnarray}
\label{eq: 4_colours}
\rm{R^{5g}:}&\rightarrow \frac{1}{\sqrt{2}}&(0,1,-1)\nonumber\\
\rm{B^{5g}:}&\rightarrow \frac{1}{\sqrt{2}}&(1,-1,0)\nonumber\\
\rm{G^{5g}:}&\rightarrow \frac{1}{\sqrt{2}}&(-1,0,1)
\end{eqnarray}

while for $\Gamma^{4g}$:

\begin{eqnarray}
\label{eq: 4_colours}
\rm{R^{4g}:}&\rightarrow \frac{1}{\sqrt{6}}&(-2,1,1)\nonumber\\
\rm{B^{4g}:}&\rightarrow \frac{1}{\sqrt{6}}&(1,1,-2)\nonumber\\
\rm{G^{4g}:}&\rightarrow \frac{1}{\sqrt{6}}&(1,-2,1)
\end{eqnarray}

By inspection, the three-fold rotations permute the three colours, the $[1,0,0]$-2-fold rotations leave all colours invariant, whilst the  $[1,0,0]$-4-fold rotations and $[1,1,0]$-2-fold rotations leave one colour invariant and permute the other two.  Hence, the colour group symbol is $\{\hmn{432}\vert \hmn{422} \vert {222}\}$.  The point group $\hmn{422}$ is a subgroup of index three, so  $\{\hmn{432}\vert \hmn{422} \vert {222}\}$ is indeed a three-colour group.

Next, we construct the coloured tensors with the projection method, starting from a generic rank-2 tensor (Eq. \ref{eq: ten_gen}) and, this time, assigning the colour red (R) to it:

\begin{eqnarray}
\label{eq: ten_colour_Mn3GaN}
\ten{T}_R&=\frac{1}{6}\left(
\begin{array}{ccc}
2a & 0 & 0  \\
0 & b+c & 0   \\
 0& 0 & b+c  \\
\end{array}
\right)\nonumber\\
\ten{T}_B&=\frac{1}{6}\left(
\begin{array}{ccc}
b+c & 0 & 0  \\
0 & b+c & 0   \\
 0& 0 & 2a  \\
\end{array}
\right)\nonumber\\
\ten{T}_G&=\frac{1}{6}\left(
\begin{array}{ccc}
b+c & 0 & 0  \\
0 & 2a & 0   \\
 0& 0 & b+c  \\
\end{array}
\right)\nonumber
\end{eqnarray}

Re-assembling the full CPG tensor we obtain for $\Gamma^{5g}$:

\begin{equation}
\label{eq: tensor_recomp_gamma5_3x6}
\ten{T}_{\rm{CPG}}^{5g}=\Lambda  \left(
\begin{array}{cccccc}
 0 & 1  & -1 & 0& 0 & 0 \\
 -1 & 0 & 1 & 0 & 0 & 0 \\
1 & -1 & 0 & 0 & 0 & 0 \\
\end{array}
\right)
\end{equation}

with $\Lambda =-(1/(6 \sqrt{2})) (2 a - b - c)$, while for $\Gamma^{4g}$:

\begin{equation}
\label{eq: tensor_recomp_gamma4_3x6}
\ten{T}_{\rm{CPG}}^{4g}=\Lambda \left(
\begin{array}{cccccc}
 2 &-1  & -1 & 0& 0 & 0 \\
 -1 & 2 & -1 & 0 & 0 & 0 \\
 -1 & -1 & 2 & 0 & 0 & 0 \\
\end{array}
\right)
\end{equation}

with $\Lambda =-(1/(6 \sqrt{6})) (2 a - b - c)$

The two tensor are orthogonal, meaning that the spin textures will be orthogonal at any wavevector.

As already explained, the MPGs for the $\Gamma^{5g}$ and $\Gamma^{4g}$ ($\hmn{-3m}$ and $\hmn{-3m'}$, Class VIII and Class VII, respectively in Table I of Ref. \onlinecite{Radaelli2024}) belong to the trigonal system.  Therefore, in order to compare them to the CPG tensors, we need to rotate them to the cubic coordinates.  The simple transformations yield:  

\begin{equation}
\label{eq: tensor_3barm}
\ten{T}^{\hmn{-3m}}=\Lambda_1 \left(
\begin{array}{cccccc}
 0 & 1 & -1 & 0&0 &0 \\
 -1& 0 & 1 & 0 & 0 &0 \\
1& -1 & 0 & 0& 0 & 0 \\
\end{array}
\right)+\Lambda_2 \left(
\begin{array}{cccccc}
 0 & 0  & 0 & 0& 1& -1 \\
 0& 0 & 0& -1 & 0 & 1\\
0& 0 & 0 & 1 & -1 & 0 \\
\end{array}
\right)
\end{equation}

\begin{equation}
\label{eq: tensor_3barmp}
\ten{T}^{\hmn{-3m'}}=\Lambda_1 \left(
\begin{array}{cccccc}
 2 &-1  & -1& 0& 0 & 0 \\
 -1 & 2 & -1 & 0 & 0 & 0 \\
 -1 & -1 & 2 & 0 & 0 & 0 \\
\end{array}
\right)+
\Lambda_2 \left(
\begin{array}{cccccc}
0& 0 & 0 & 2 &-1  & -1  \\
 0 & 0 & 0 &-1 & 2 & -1\\
 0 & 0 & 0 & -1 & -1 & 2\\
\end{array}
\right)+
\Lambda_3 \left(
\begin{array}{cccccc}
 1 &1  & 1& 0& 0 & 0 \\
 1 & 1 & 1 & 0 & 0 & 0 \\
1 & 1 & 1 & 0 & 0 & 0 \\
\end{array}
\right)
\end{equation}

One can easily see that $\ten{T}_{\rm{CPG}}^{5g}$  (Eq. \ref {eq: tensor_recomp_gamma5_3x6}) is equal to the first term of $\ten{T}^{\hmn{-3m}}$ (Eq. \ref{eq: tensor_3barm} ) with $\Lambda \rightarrow \Lambda_1$.  Likewise, $\ten{T}_{\rm{CPG}}^{4g}$  (Eq. \ref {eq: tensor_recomp_gamma4_3x6}) is equal to the first term of $\ten{T}^{\hmn{-3m'}}$ (Eq. \ref{eq: tensor_3barmp} ) with $\Lambda \rightarrow \Lambda_1$.  In other words, tensors obtained from the CPG analysis are \emph{special cases} of the general tensors allowed by the MPG, obtained, in this case by setting $\Lambda_2=\Lambda_3=0$.  This is exactly what we expected and is completely consistent with the result we previously obtained for collinear structures (Ref. \onlinecite{Radaelli2024}).

\begin{figure}[!h]
\centering
\includegraphics[scale=0.5]{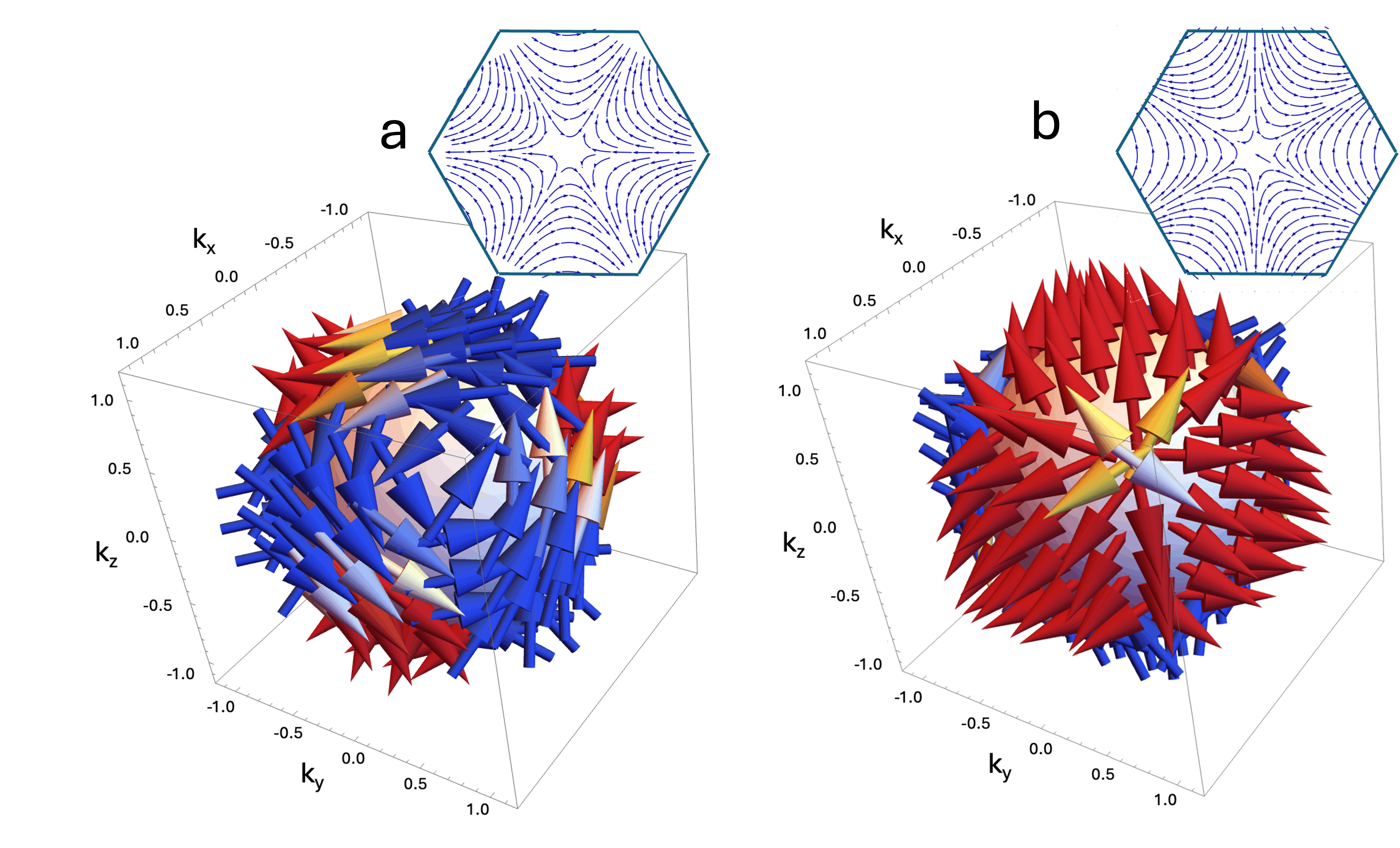}
\caption{ \label{fig: Mn3GaN_texture}  (Colour online) Altermagnetic-like spin textures for the $\Gamma^{5g}$ (a) $\Gamma^{4g}$ (b) and phases of Mn$_3$GaN, generated from the tensors $\ten{T}_{\rm{CPG}}^{5g}$ and $\ten{T}_{\rm{CPG}}^{4g}$  (Eqs. \ref {eq: tensor_recomp_gamma5_3x6} and \ref{eq: tensor_recomp_gamma4_3x6}).  The colour of the arrows indicates the radial projection of the texture (red=out; yellow/grey=parallel; blue=in). Wavevectors are in arbitrary units. The insets are the equatorial cross sections of the textures down the (1,1,1) direction, shown with an outline of the Brillouin zone (compare with Fig. \ref{fig: Mn3GaN_DFT} ).  The $\Gamma^{5g}$ and $\Gamma^{4g}$texture are related by a collective 90$^\circ$-rotation in spin space around the (1,1,1) direction.}
\end{figure}

\subsection{M\lowercase{n}$_3$G\lowercase{a}N and M\lowercase{n}$_3$I\lowercase{r}(G\lowercase{e},S\lowercase{i}): SG analysis}
\label{sec: SG CPG parallel}

Here, we give examples of analyses performed on Mn$_3$GaN and  Mn$_3$Ir(Ge,Si) with the SG approach and using the terminology and numbering of Ref. \onlinecite{Litvin1977}.   As already mentioned, the SGs corresponding to a given CPG are obtained from the XPG $G (R)$ and one of its normal subgroups $H (r)$, where the letter in brackets corresponds to the notation of Ref. \onlinecite{Litvin1977}. 

In the case of Mn$_3$GaN, $G=\hmn{432}$ and $H=\hmn{222}$.  The spin group is constructed by selecting spin-transformation PGs (denoted as $B$) that are isomorphic to $G/H$ (or $R/r$ in the notation of Ref. \onlinecite{Litvin1977}).  There are two possible PGs for $B$: $\hmn{32}$ (SG no. 563) and $\hmn{3m}$ (SG no. 564), which are, indeed, both isomorphic to the three-colour permutation group.  In Ref. \onlinecite{Litvin1977}, it is explained that the two coordinate systems for $G$ and $B$ are arbitrarily mutually orientated, though a complex notation is introduced when the coordinate systems of $B$ and $G$ do not coincide.  However, importantly, one needs to orient the axes of $B$ so that they act correctly on the spin system in real space.  In our case, the three-fold axis must be perpendicular to the three spins, each two-fold axis in $32$ must be parallel to one of the spins, and each mirror plane in $3m$ must be perpendicular to one of the spins.  When this is done, the analysis is identical to the one using CPGs, since $B$ acts by permuting the spins in the same way as colours were permuted by the CPG.  Note, however, the $B$ needs to be rotated by 90$^\circ$ rotation between $\Gamma^{5g}$ and $\Gamma^{4g}$ and more generally re-oriented to fit an arbitrary rotation in spin space.  Leaving aside the  $\hmn{32}/\hmn{3m}$ ambiguity, this seems an unnecessary complication that is completely avoided by the CPG analysis.  Within the tensorial framework,  the CPG analysis also captures the fundamentally \emph{scalar} nature of altermagnetic-like tensors.

Similarly, a parallel SG analysis can be performed on Mn$_3$Ir(Ge,Si), where $G=\hmn{23}$ and $H=\hmn{1}$ (SG 538 in Ref. \onlinecite{Litvin1977}). Here, the SG is the \emph{same} for the approximate and experimental magnetic structures (see Sec. \ref{sec: 4_colours} and \ref{sec: 12_colours}), while the CPG are different --- $\{23 \vert 3 \vert 1 \}$ vs $\{23 \vert 1 \vert 1 \}$ because, as already explained, the CPG symbol encode additional information about the number of colours (i.e., distinct spin directions) and the specific representation that encodes the colour permutation.

\subsection{M\lowercase{n}$_3$G\lowercase{a}N: DFT calculations and tensor decompositions}

To validate our analysis further, we calculated the spin textures for the $\Gamma^{5g}$ and $\Gamma^{4g}$ phases of Mn$_3$GaN using spin-resolved DFT. DFT calculations were performed using the Quantum-ESPRESSO code\cite{QE-2017}. The exchange and correlation effects were treated within the generalized gradient approximation (GGA)\cite{GGA}. The k-point mesh of $16\times16\times16$ and plane-wave cut-off energy of $52$ Ry were used for the integration in the irreducible Brillouin zone. The non-collinear magnetic structure was always considered with and without spin orbit coupling. The spin texture on the Fermi surface was calculated using $50\times50\times50$ k points within the first Brillouin zone (BZ). To depict the results in an intuitive manner, the spin texture was projected on a plane containing the $\Gamma$ point and normal to the $(1,1,1)$ direction.

\begin{figure}[!h]
\centering
\includegraphics[scale=0.75]{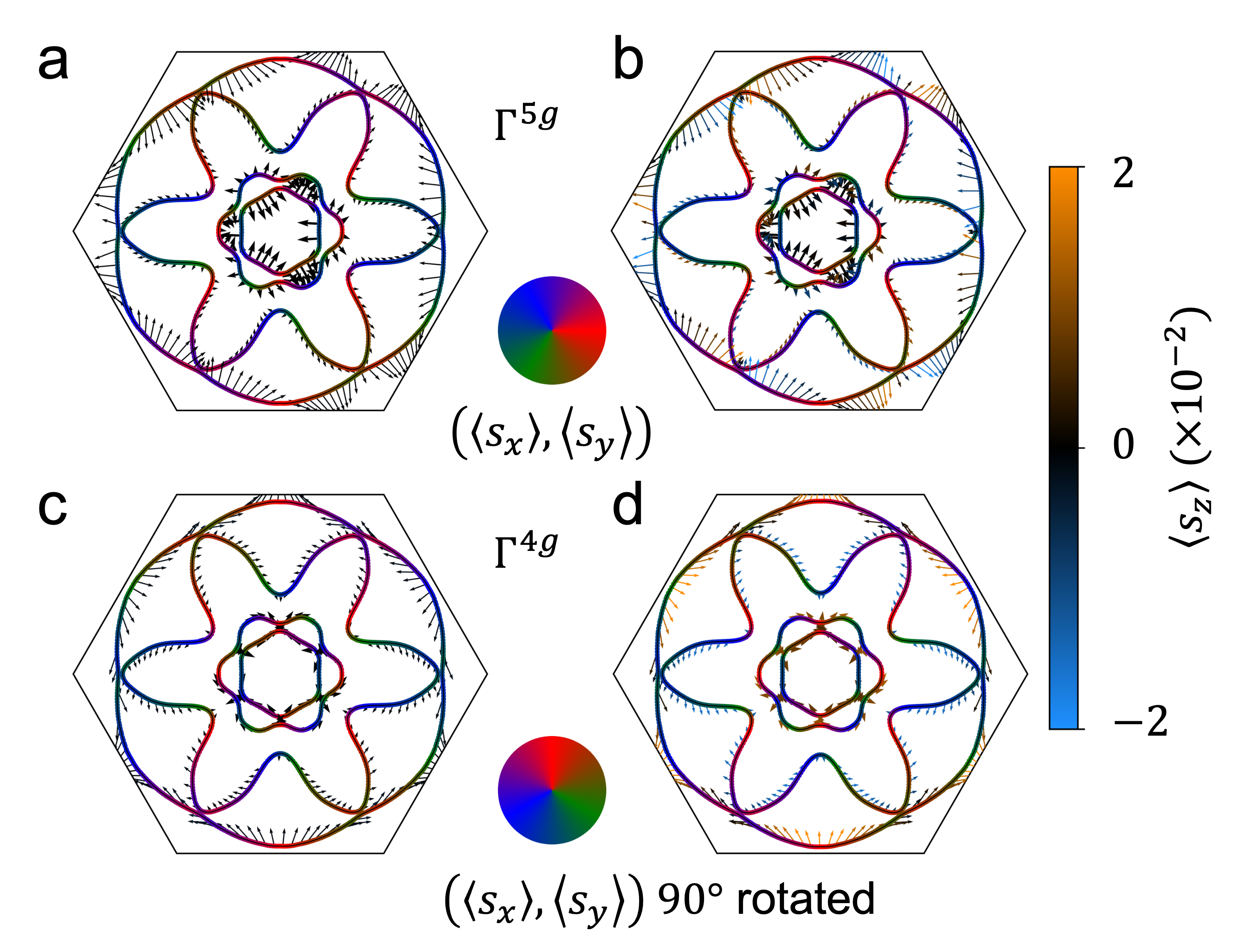}
\caption{ \label{fig: Mn3GaN_DFT} (Colour online) Spin texture at the Fermi surface on a plane cut perpendicularly to the $(1,1,1)$ direction including the $\Gamma$ point shown with an outline of the BZ, calculated using spin-resolved DFT for the $\Gamma^{5g}$ (top) and  $\Gamma^{4g}$ (bottom) phases with (a, c) and without (b,d) SOC.  For $\Gamma^{4g}$, the colour wheel indicating in-plane spin directions was rotated by $90^\circ$.  The $z$-component, which is only present with SOC, is also indicated.  For $\Gamma^{4g}$, each band has a net magnetisation. The $z$ direction is along $(1,1,1)$.}
\end{figure}

Fig. \ref{fig: Mn3GaN_DFT} displays \textcolor{\altcolor} {equatorial sections (i.e., on a plane cut perpendicularly to the (1, 1, 1) direction)} of the spin-resolved Fermi surface, calculated for $\Gamma^{5g}$ and $\Gamma^{4g}$ with and without SOC.  For $\Gamma^{4g}$, the colour wheel indicating in-plane spin directions was rotated by $90^\circ$, to emphasise the expected relation with the $\Gamma^{5g}$ texture.  Band dispersions along high-symmetry directions are displayed in Fig. S2 and Fig. S3 of the Supplement Information\cite{supp}.  In the absence of SOC, the  $\Gamma^{5g}$ (panel a) and $\Gamma^{4g}$ (panel c) are identical except for the $90^\circ$ rotation, and there is no spin-texture component along $(1,1,1)$, as predicted by the CPG analysis at any tensor rank.  The spin textures are also very similar to those in Fig. \ref{fig: Mn3GaN_texture} (see insets for the same cuts).   By contrast, in the presence of SOC, the $\Gamma^{5g}$ (panel b) and $\Gamma^{4g}$ (panel d) become significantly different, and acquire a component along $(1,1,1)$.  In the case of $\Gamma^{5g}$, this component arises from a higher-rank tensor \textcolor{\altcolor}{(see below)} and averages to zero for each band, while for $\Gamma^{4g}$ the additional $\Lambda_3$ term (Eq. \ref{eq: tensor_3barmp}) yields a net spin polarisation in each band, and an overall weak magnetisation for the whole structure.  


\begin{figure}[!h]
\centering
\includegraphics[scale=0.6]{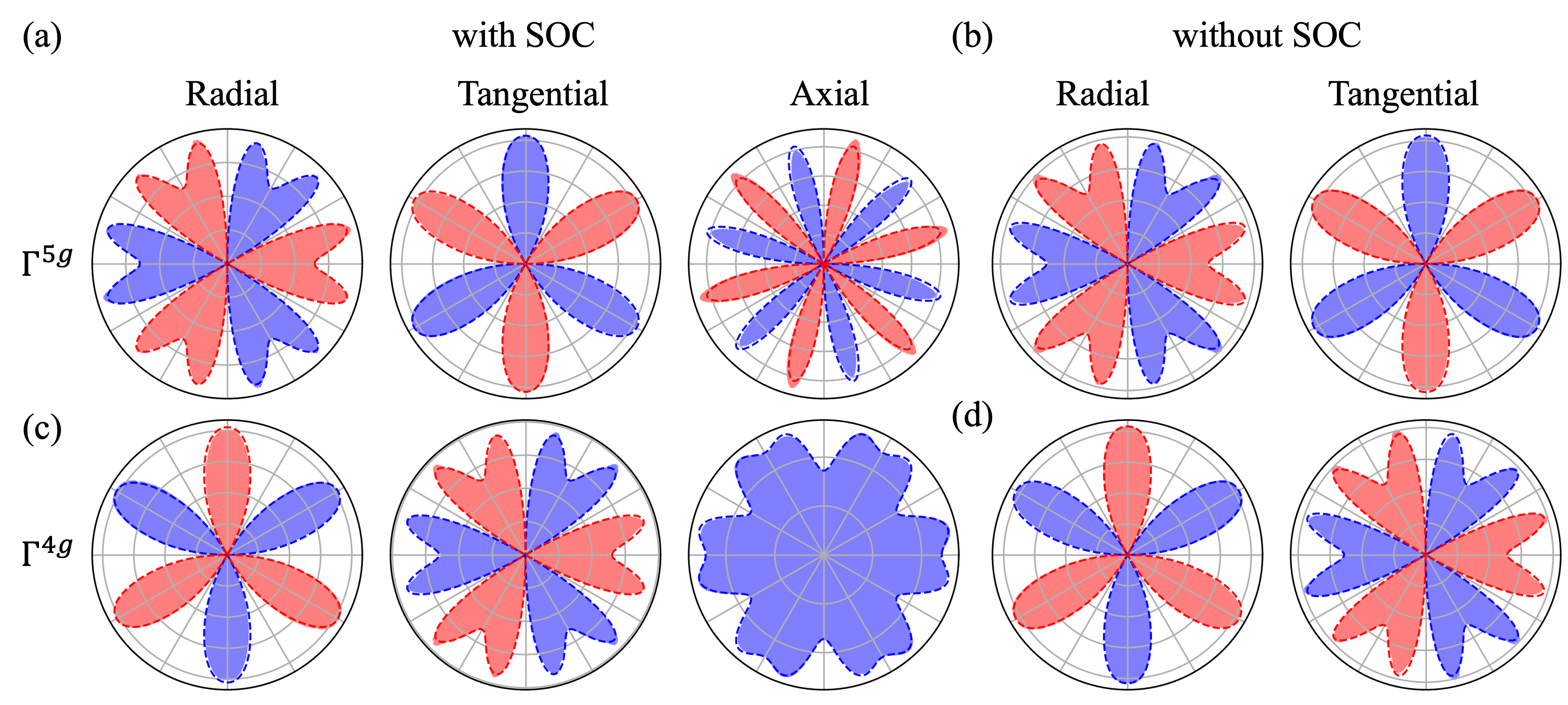}
\caption{ \label{fig: Mn3GaN_Band1} (Colour online) Decomposition of the Band-1 spin textures calculated with DFT (solid blocks) and with the corresponding tensorial fits (dashed lines) for $\Gamma^{5g}$ (top row) and $\Gamma^{4g}$ (bottom row).  Red/Blue indicate positive/negative components (note the $\vec{k}$/$-\vec{k}$ symmetry). Panels (a)/(c) and (b)/(d) were calculated, respectively with and without SOC (see text). \textcolor{\altcolor}{The outer circles in the plots corresponds to amplitudes for the radial, tangential and axial components of [0.190,0.208,0.009]/[0.205,0.190,0.013] or (0.198,0.208)/(0.208,0.198) for the $\Gamma^{5g}$/$\Gamma^{4g}$ phase, respectively, with $[]$ or without $()$ SOC respectively.}}
\end{figure}

\textcolor{\altcolor}{To demonstrate the effectiveness of our tensorial approach, we fitted the spin textures from DFT onto a tensorial expansions based on CPG (no SOC) and MPG (with SOC) analyses for both $\Gamma^{4g}$ and $\Gamma^{5g}$.  To this end, we decomposed the spin textures of the equatorial sections (i.e., sections perpendicular to $(111)$, as in Fig. \ref{fig: Mn3GaN_DFT}) for the four bands that cross the Fermi surface into a \emph{radial} component perpendicular to $\vec{k}$ and to $(111)$, a \emph{tangential} component in the equatorial plane and perpendicular to $\vec{k}$, and an \emph{axial} component parallel to $(111)$, which is zero in the absence of SOC.  (\textcolor{\altcolor}{The bands are labelled from 0 to 3 as we move away from the $\Gamma$ point as shown in the SI, Supplementary Fig.S4})  Note that, with DFT, the spin texture is calculated on constant-energy surfaces rather than at constant $|\vec{k}|$.  However, since constant-energy surfaces have at least the full symmetry of the XPG, textures will have an identical tensorial expansion, though the coefficients will be different.  CPG tensor decompositions were performed up to tensorial rank 18, while MPG decompositions of the axial component were to rank 13 (we recall that MPG tensors are odd-ranked, while CPG tensors are even-ranked).  The three tensorial components of the spin texture are linear combinations of sine and cosine functions of the azimuthal angle $\phi$ in the equatorial plane.  The coefficients of this expansion (four for the radial and tangential components, three for the axial component) were fitted to the corresponding components of the DFT spin textures.  The results of these fits are displayed graphically in Fig. \ref{fig: Mn3GaN_Band1} for Band 1.  Corresponding representations for the other bands are in Fig. S5-S7}, while  the functional form and fitted parameters are displayed in Table S1 and S2. 

\textcolor{\altcolor}{In the absence of SOC, the radial and tangential components are swapped between  $\Gamma^{5g}$ and $\Gamma^{4g}$, while the axial component is absent.  This is completely consistent with the CG analysis, and, in particular, with the fact that texture for $\Gamma^{5g}$ and $\Gamma^{4g}$ are rotated by 90$^\circ$ in spin space. When SOC is switched on, a small axial component (\textcolor{\altcolor}{approximately a factor of 10 smaller}) appears for both $\Gamma^{5g}$ and $\Gamma^{4g}$.  However, for $\Gamma^{5g}$ the axial spin polarisation averaged over the Fermi surface is zero, while for $\Gamma^{4g}$ there is a net spin polarisation, consistent with weak ferromagnetism.  Further details of the effects of SOC on the spin-polarised band structures are evident from the band dispersions in Fig S3 and S2 in the SI. Particularly noteworthy is the fact that the bands along the $\Gamma-R$ direction must, by symmetry, have polarisation along the $(111)$ direction.  Hence, a small band splitting is observed only for $\Gamma^{4g}$ in the presence of SOC, where the band remains unsplit in all other cases (see insets).}

\section{Discussion and conclusions}
\label{sec: discussion}

In this paper, `altermagnetic-like' spin textures have been defined as the components of the momentum-space spin texture that is invariant by global rotations in spin space in the sense of Eq. \ref{eq: invariance}. The tensorial forms of these textures can be defined consistently based on CPG analysis for both non-collinear and collinear AFM structures, since the latter are a special example of bi-colour (black and white) symmetry.   Strong FM textures (i.e., those present in a true ferro/ferrimagnet rather than a weak/canted FM) are also a special example, for which the colour group is trivial (colourless).\footnote{Note that the net magnetization in the absence of SOC is \emph{identically zero} for collinear altermagnets and also for non-collinear magnets where each colour has a corresponding anti-colour, since the magnetisation of anti-coloured sublattices cancels.}  In addition to altermagnetic-like textures, the MPG of the crystal/magnetic structure will generally allow additional ('non-altermagnetic') spin textures in momentum space, which will be specific to a given orientation of the spin system and will be absent in the absence of SOC. Altermagnetic-like/ferromagnetic textures are allowed at some tensorial order for all MPGs that do not explicitly forbids $\vec{k}/-\vec{k}$ \emph{symmetric}, time-reversal odd spin textures (i.e., those that do not contain the time reversal operator or the product of inversion and time reversal operators).  The MPG/CPG pair defines which part of the full MPG texture (tabulated to lowest tensorial order in Ref. \onlinecite{Radaelli2024}) is altermagnetic-like, with the CPG also providing the `tensorial building blocks' (coloured tensors) to reconstruct the full altermagnetic-like texture, once the exact real-space spin orientations are defined.  These building blocks can in principle be derived for all CPGs once and for all, though calculating them for a specific magnetic structure is a simple exercise.   We have demonstrated the CPG approach for three non-collinear magnetic structures --- Mn$_3$Ir(Ge,Si), Pb$_2$MnO$_4$ and  Mn$_3$GaN.  In the case of  Mn$_3$GaN, using DFT we have also calculated the spin textures for two different magnetic phases ($\Gamma^{5g}$ and $\Gamma^{4g}$) with and without SOC, and demonstrated that they are entirely consistent with the predictions of MPG/CPG theory.  \textcolor{\altcolor}{We also find that the tensorial decomposition of the DFT textures is in very good agreement with CPG (no SOC) and MPG (with SOC) predictions, confirming the soundness of this method.}

Beyond the results reported in this paper, we remark that CPG/CSG symmetry can also be defined for $p$-wave magnets\cite{Hellenes2023} and other systems (e.g., based on triangular lattices),\cite{Hayami2020b} which allow $\vec{k}/-\vec{k}$ \emph{anti-symmetric} , time-reversal even spin textures in momentum space, and it is expected that those spin textures will also be constrained by CSG symmetry.  However, our tensorial analysis will need to be modified to deal with such systems.\footnote{Very briefly, magnets with $\vec{k}/-\vec{k}$ \emph{anti-symmetric} splitting are characterised by bond-type multipolar ordering,\cite{Hayami2020b} and the staggered magnetisation (in the collinear case) or the one-colour axial unit vectors (non-collinear case) need to be replaced in the tensorial treatment by the cross product $\bm{S}_1 \times \bm{S}_2$, where sites 1 and 2 are those connected by the specific bond.  This approach will be fully described in a future paper.}

%
\appendix

\section{Colour symmetry, representations and exchange multiplets}
\label{sec: Izyumov}

There is a close relation between the description of magnetic structures in terms of CGs and the theory of exchange multiplets, \cite{Izyumov1980} which classifies magnetic structures in terms of irreducible representations (\textit{irreps}) of the symmetry group of a Hamiltonian that only involves symmetric exchange and is therefore invariant by rotation in spin space.  The symmetry of this Hamiltonian is higher than the crystal symmetry, thus, in general, several \textit{irreps} of the space group will be combined into so-called `exchange multiplets', and may be activated simultaneously at a magnetic phase transition.  Exchange multiplets are constructed as tensor products of a \emph{single} \textit{irrep} of the space group that is contained in the full permutational (scalar) representation of the equivalent magnetic sites (with dimensionality equal to the number of sites), times the (generally reducible) axial-vector representation  (see Ref. \onlinecite{Izyumov1980} ).  

By contrast, the set of all possible magnetic structures with a given CG and for a given set of equivalent magnetic sites is constructed as the tensor product of the `colour-permuting representation' times the axial-vector representation.  In turn, the colour-permuting representation is the permutational representation of the magnetic sites restricted to the subspace defined by a one-colour basis set, where all the sites with a given colour are assigned the number `1', while all the other sites are assigned `0'. This basis set can be easily constructed projectively from the one-site basis of the permutational representation.  This is done from the left-coset decomposition of $H'$ in $G$:

\begin{equation}
\label{eq: coset}
G=H'+g_1 \circ H' + g_2 \circ H' + \dots
\end{equation}

Let $\bm{b}$ be an element of the one-site basis set of the permutational representation of the magnetic sites, such that one of the sites is associated with the number `1' and all the others with '0'.  Let $h'_i,\; i=1\dots n$ be all the elements of $H'$.  Then the elements of the one-colour basis set, say $\bm{b}_R$, $\bm{b}_Y$, $\bm{b}_G$, etc., are defined as

\begin{eqnarray}
\bm{b}_R&=&\frac{1}{n} \sum_{i=1}^n h'_i [\bm{b}]\nonumber\\
\bm{b}_Y&=&\frac{1}{n} \sum_{i=1}^n g_1\circ h'_i [\bm{b}]\nonumber\\
\bm{b}_G&=&\frac{1}{n} \sum_{i=1}^n g_2\circ h'_i [\bm{b}]\nonumber\\
&&\dots
\end{eqnarray}

One can easily see that this basis is in accord with the definition of the CGs, whereby `red' sites are those where $\bm{b}_R$ is non-zero, etc. $H'$ brings red sites into red sites, $g_1\circ H'$ brings red sites into yellow sites, etc.  The CG also defines the representation of $G$ onto the linear space defined by the one-colour basis set, since the transformation matrices can be written out explicitly. 

A basis set for the linear space of all magnetic structures with a given colour group is:

\begin{equation}
\label{eq: tensor_basis_set}
 \left(
 \bm{b}_R \vec{\hat{i}},\bm{b}_R \vec{\hat{j}}, \bm{b}_R \vec{\hat{k}}, 
 \bm{b}_Y \vec{\hat{i}},\bm{b}_Y \vec{\hat{j}}, \bm{b}_Y \vec{\hat{k}},
 \bm{b}_G \vec{\hat{i}},\bm{b}_G \vec{\hat{j}}, \bm{b}_G \vec{\hat{k}}, \dots \right)
\end{equation}

A generic elements of this linear space is:

\begin{equation}
\bm{b}_R \, \vec{v}_1 + \bm{b}_Y \, \vec{v}_2 +\bm{b}_G \, \vec{v}_3 +\dots
\end{equation}

where $\vec{v}_1$, $\vec{v}_2$, etc. are generic linear combinations of the unit vectors $\vec{\hat{i}}$, $\vec{\hat{j}}$ and $\vec{\hat{k}}$.  This corresponds to assign a generic vector to each of the colours, exactly as we anticipated.

Finally, since the colour-permuting representation, with dimensionality equal to the number of colours, is generally \emph{reducible}, it follows that each CG describes several exchange multiplets.  

\begin{acknowledgements}
The authors would like to acknowledge the use of the University of Oxford Advanced Research Computing (ARC) facility (\url {http://dx.doi.org/10.5281/zenodo.22558}) in carrying out this work.
\end{acknowledgements}


\bibliography{Altermag_Tensor_2024.bib}

\begin{thebibliography}{39}%
\makeatletter
\providecommand \@ifxundefined [1]{%
 \@ifx{#1\undefined}
}%
\providecommand \@ifnum [1]{%
 \ifnum #1\expandafter \@firstoftwo
 \else \expandafter \@secondoftwo
 \fi
}%
\providecommand \@ifx [1]{%
 \ifx #1\expandafter \@firstoftwo
 \else \expandafter \@secondoftwo
 \fi
}%
\providecommand \natexlab [1]{#1}%
\providecommand \enquote  [1]{``#1''}%
\providecommand \bibnamefont  [1]{#1}%
\providecommand \bibfnamefont [1]{#1}%
\providecommand \citenamefont [1]{#1}%
\providecommand \href@noop [0]{\@secondoftwo}%
\providecommand \href [0]{\begingroup \@sanitize@url \@href}%
\providecommand \@href[1]{\@@startlink{#1}\@@href}%
\providecommand \@@href[1]{\endgroup#1\@@endlink}%
\providecommand \@sanitize@url [0]{\catcode `\\12\catcode `\$12\catcode `\&12\catcode `\#12\catcode `\^12\catcode `\_12\catcode `\%12\relax}%
\providecommand \@@startlink[1]{}%
\providecommand \@@endlink[0]{}%
\providecommand \url  [0]{\begingroup\@sanitize@url \@url }%
\providecommand \@url [1]{\endgroup\@href {#1}{\urlprefix }}%
\providecommand \urlprefix  [0]{URL }%
\providecommand \Eprint [0]{\href }%
\providecommand \doibase [0]{http://dx.doi.org/}%
\providecommand \selectlanguage [0]{\@gobble}%
\providecommand \bibinfo  [0]{\@secondoftwo}%
\providecommand \bibfield  [0]{\@secondoftwo}%
\providecommand \translation [1]{[#1]}%
\providecommand \BibitemOpen [0]{}%
\providecommand \bibitemStop [0]{}%
\providecommand \bibitemNoStop [0]{.\EOS\space}%
\providecommand \EOS [0]{\spacefactor3000\relax}%
\providecommand \BibitemShut  [1]{\csname bibitem#1\endcsname}%
\let\auto@bib@innerbib\@empty
\bibitem [{\citenamefont {{\v{S}}mejkal}\ \emph {et~al.}(2022{\natexlab{a}})\citenamefont {{\v{S}}mejkal}, \citenamefont {Sinova},\ and\ \citenamefont {Jungwirth}}]{Smejkal2022}%
  \BibitemOpen
  \bibfield  {author} {\bibinfo {author} {\bibfnamefont {L.}~\bibnamefont {{\v{S}}mejkal}}, \bibinfo {author} {\bibfnamefont {J.}~\bibnamefont {Sinova}}, \ and\ \bibinfo {author} {\bibfnamefont {T.}~\bibnamefont {Jungwirth}},\ }\href {\doibase 10.1103/PHYSREVX.12.031042/FIGURES/4/MEDIUM} {\bibfield  {journal} {\bibinfo  {journal} {Physical Review X}\ }\textbf {\bibinfo {volume} {12}},\ \bibinfo {pages} {031042} (\bibinfo {year} {2022}{\natexlab{a}})}\BibitemShut {NoStop}%
\bibitem [{\citenamefont {{\v{S}}mejkal}\ \emph {et~al.}(2022{\natexlab{b}})\citenamefont {{\v{S}}mejkal}, \citenamefont {Sinova},\ and\ \citenamefont {Jungwirth}}]{Smejkal2022b}%
  \BibitemOpen
  \bibfield  {author} {\bibinfo {author} {\bibfnamefont {L.}~\bibnamefont {{\v{S}}mejkal}}, \bibinfo {author} {\bibfnamefont {J.}~\bibnamefont {Sinova}}, \ and\ \bibinfo {author} {\bibfnamefont {T.}~\bibnamefont {Jungwirth}},\ }\href {\doibase 10.1103/PHYSREVX.12.040501/FIGURES/14/MEDIUM} {\bibfield  {journal} {\bibinfo  {journal} {Physical Review X}\ }\textbf {\bibinfo {volume} {12}},\ \bibinfo {pages} {040501} (\bibinfo {year} {2022}{\natexlab{b}})},\ \Eprint {http://arxiv.org/abs/2204.10844} {arXiv:2204.10844} \BibitemShut {NoStop}%
\bibitem [{\citenamefont {Radaelli}(2024)}]{Radaelli2024}%
  \BibitemOpen
  \bibfield  {author} {\bibinfo {author} {\bibfnamefont {P.~G.}\ \bibnamefont {Radaelli}},\ }\href {\doibase 10.1103/PhysRevB.110.214428} {\bibfield  {journal} {\bibinfo  {journal} {Physical Review B}\ }\textbf {\bibinfo {volume} {110}},\ \bibinfo {pages} {214428} (\bibinfo {year} {2024})}\BibitemShut {NoStop}%
\bibitem [{\citenamefont {Cheong}\ and\ \citenamefont {Huang}(2024)}]{Cheong2024}%
  \BibitemOpen
  \bibfield  {author} {\bibinfo {author} {\bibfnamefont {S.~W.}\ \bibnamefont {Cheong}}\ and\ \bibinfo {author} {\bibfnamefont {F.~T.}\ \bibnamefont {Huang}},\ }\href {\doibase 10.1038/s41535-024-00626-6} {\bibfield  {journal} {\bibinfo  {journal} {npj Quantum Materials 2024 9:1}\ }\textbf {\bibinfo {volume} {9}},\ \bibinfo {pages} {1} (\bibinfo {year} {2024})}\BibitemShut {NoStop}%
\bibitem [{\citenamefont {Hellenes}\ \emph {et~al.}(2023)\citenamefont {Hellenes}, \citenamefont {Jungwirth}, \citenamefont {Jaeschke-Ubiergo}, \citenamefont {Chakraborty}, \citenamefont {Sinova},\ and\ \citenamefont {{\v{S}}mejkal}}]{Hellenes2023}%
  \BibitemOpen
  \bibfield  {author} {\bibinfo {author} {\bibfnamefont {A.~B.}\ \bibnamefont {Hellenes}}, \bibinfo {author} {\bibfnamefont {T.}~\bibnamefont {Jungwirth}}, \bibinfo {author} {\bibfnamefont {R.}~\bibnamefont {Jaeschke-Ubiergo}}, \bibinfo {author} {\bibfnamefont {A.}~\bibnamefont {Chakraborty}}, \bibinfo {author} {\bibfnamefont {J.}~\bibnamefont {Sinova}}, \ and\ \bibinfo {author} {\bibfnamefont {L.}~\bibnamefont {{\v{S}}mejkal}},\ }\href {https://arxiv.org/abs/2309.01607v3} {\  (\bibinfo {year} {2023})},\ \Eprint {http://arxiv.org/abs/2309.01607} {arXiv:2309.01607} \BibitemShut {NoStop}%
\bibitem [{\citenamefont {Hayami}\ \emph {et~al.}(2040)\citenamefont {Hayami}, \citenamefont {Yanagi},\ and\ \citenamefont {Kusunose}}]{Hayami2020b}%
  \BibitemOpen
  \bibfield  {author} {\bibinfo {author} {\bibfnamefont {S.}~\bibnamefont {Hayami}}, \bibinfo {author} {\bibfnamefont {Y.}~\bibnamefont {Yanagi}}, \ and\ \bibinfo {author} {\bibfnamefont {H.}~\bibnamefont {Kusunose}},\ }\href {\doibase 10.1103/PhysRevB.101.220403} {\bibfield  {journal} {\bibinfo  {journal} {Physical Review B}\ }\textbf {\bibinfo {volume} {101}},\ \bibinfo {pages} {220403} (\bibinfo {year} {2040})}\BibitemShut {NoStop}%
\bibitem [{Note1()}]{Note1}%
  \BibitemOpen
  \bibinfo {note} {`$\protect \mathbf {k}/-\protect \mathbf {k}$ \protect \emph {symmetric}/\protect \emph {anti-symmetric}' means that the spin texture, defined as $ \protect \mathbf {s}_{n\protect \mathbf {k}}=\mathinner {\langle {\Psi _{n\protect \mathbf {k}}}|}\protect \bm {\sigma }\mathinner {|{\Psi _{n\protect \mathbf {k}}}\rangle }$ (see below) is the same/changes sign by exchanging $\protect \mathbf {k}$ and $-\protect \mathbf {k}$. `Time-reversal odd/even' means that the spin texture in the time-reversed magnetic domain (i.e., the domain obtained by flipping all the magnetic moments) is opposite/the same for a given $\protect \mathbf {k}$.}\BibitemShut {Stop}%
\bibitem [{\citenamefont {Dresselhaus}(1955)}]{Dresselhaus1955}%
  \BibitemOpen
  \bibfield  {author} {\bibinfo {author} {\bibfnamefont {G.}~\bibnamefont {Dresselhaus}},\ }\href {\doibase 10.1103/PhysRev.100.580} {\bibfield  {journal} {\bibinfo  {journal} {Physical Review}\ }\textbf {\bibinfo {volume} {100}},\ \bibinfo {pages} {580} (\bibinfo {year} {1955})}\BibitemShut {NoStop}%
\bibitem [{\citenamefont {Rashba}\ and\ \citenamefont {Sheka}(1959)}]{Rashba1959}%
  \BibitemOpen
  \bibfield  {author} {\bibinfo {author} {\bibfnamefont {E.~I.}\ \bibnamefont {Rashba}}\ and\ \bibinfo {author} {\bibfnamefont {V.}~\bibnamefont {Sheka}},\ }\href@noop {} {\bibfield  {journal} {\bibinfo  {journal} {Fiz. Tverd. Tela: Collected Papers}\ }\textbf {\bibinfo {volume} {2}},\ \bibinfo {pages} {62} (\bibinfo {year} {1959})}\BibitemShut {NoStop}%
\bibitem [{\citenamefont {{\v{Z}}elezn{\'{y}}}\ \emph {et~al.}(2017)\citenamefont {{\v{Z}}elezn{\'{y}}}, \citenamefont {Zhang}, \citenamefont {Felser},\ and\ \citenamefont {Yan}}]{Zelezny2017}%
  \BibitemOpen
  \bibfield  {author} {\bibinfo {author} {\bibfnamefont {J.}~\bibnamefont {{\v{Z}}elezn{\'{y}}}}, \bibinfo {author} {\bibfnamefont {Y.}~\bibnamefont {Zhang}}, \bibinfo {author} {\bibfnamefont {C.}~\bibnamefont {Felser}}, \ and\ \bibinfo {author} {\bibfnamefont {B.}~\bibnamefont {Yan}},\ }\href {\doibase 10.1103/PHYSREVLETT.119.187204/FIGURES/3/MEDIUM} {\bibfield  {journal} {\bibinfo  {journal} {Physical Review Letters}\ }\textbf {\bibinfo {volume} {119}},\ \bibinfo {pages} {187204} (\bibinfo {year} {2017})},\ \Eprint {http://arxiv.org/abs/1702.00295} {arXiv:1702.00295} \BibitemShut {NoStop}%
\bibitem [{\citenamefont {Sticht}\ \emph {et~al.}(1989)\citenamefont {Sticht}, \citenamefont {H{\"{o}}ck},\ and\ \citenamefont {K{\"{u}}bler}}]{Sticht1989}%
  \BibitemOpen
  \bibfield  {author} {\bibinfo {author} {\bibfnamefont {J.}~\bibnamefont {Sticht}}, \bibinfo {author} {\bibfnamefont {K.~H.}\ \bibnamefont {H{\"{o}}ck}}, \ and\ \bibinfo {author} {\bibfnamefont {J.}~\bibnamefont {K{\"{u}}bler}},\ }\href {\doibase 10.1088/0953-8984/1/43/016} {\bibfield  {journal} {\bibinfo  {journal} {Journal of Physics: Condensed Matter}\ }\textbf {\bibinfo {volume} {1}},\ \bibinfo {pages} {8155} (\bibinfo {year} {1989})}\BibitemShut {NoStop}%
\bibitem [{Note2()}]{Note2}%
  \BibitemOpen
  \bibinfo {note} {As explained in the remainder, the CG approach can be employed to describe the magnetic structures of $\protect \mathbf {k}/-\protect \mathbf {k}$-antisymmetric magnets, but coloured tensors as defined here would require some modifications. For this reason, we will exclude$\protect \mathbf {k}/-\protect \mathbf {k}$-antisymmetric magnets from our treatment.}\BibitemShut {Stop}%
\bibitem [{\citenamefont {Brekke}\ \emph {et~al.}(2024)\citenamefont {Brekke}, \citenamefont {Sukhachov}, \citenamefont {Giil}, \citenamefont {Brataas},\ and\ \citenamefont {Linder}}]{Brekke2024}%
  \BibitemOpen
  \bibfield  {author} {\bibinfo {author} {\bibfnamefont {B.}~\bibnamefont {Brekke}}, \bibinfo {author} {\bibfnamefont {P.}~\bibnamefont {Sukhachov}}, \bibinfo {author} {\bibfnamefont {H.~G.}\ \bibnamefont {Giil}}, \bibinfo {author} {\bibfnamefont {A.}~\bibnamefont {Brataas}}, \ and\ \bibinfo {author} {\bibfnamefont {J.}~\bibnamefont {Linder}},\ }\href {https://arxiv.org/abs/2405.15823v1} {\  (\bibinfo {year} {2024})},\ \Eprint {http://arxiv.org/abs/2405.15823} {arXiv:2405.15823} \BibitemShut {NoStop}%
\bibitem [{\citenamefont {Harker}(1981)}]{Harker1981}%
  \BibitemOpen
  \bibfield  {author} {\bibinfo {author} {\bibfnamefont {D.}~\bibnamefont {Harker}},\ }\href {\doibase 10.1107/S0567739481000697} {\bibfield  {journal} {\bibinfo  {journal} {Acta Crystallographica Section A}\ }\textbf {\bibinfo {volume} {37}},\ \bibinfo {pages} {286} (\bibinfo {year} {1981})}\BibitemShut {NoStop}%
\bibitem [{\citenamefont {Sivardi{\`{e}}re}(1984)}]{Sivardiere1984}%
  \BibitemOpen
  \bibfield  {author} {\bibinfo {author} {\bibfnamefont {J.}~\bibnamefont {Sivardi{\`{e}}re}},\ }\href {\doibase 10.1107/S0108767384001197} {\bibfield  {journal} {\bibinfo  {journal} {Acta Crystallographica Section A}\ }\textbf {\bibinfo {volume} {40}},\ \bibinfo {pages} {573} (\bibinfo {year} {1984})}\BibitemShut {NoStop}%
\bibitem [{\citenamefont {Roth}(1985)}]{Roth1985}%
  \BibitemOpen
  \bibfield  {author} {\bibinfo {author} {\bibfnamefont {R.~L.}\ \bibnamefont {Roth}},\ }\href {\doibase 10.1107/S0108767385001039} {\bibfield  {journal} {\bibinfo  {journal} {Acta Crystallographica Section A}\ }\textbf {\bibinfo {volume} {41}},\ \bibinfo {pages} {484} (\bibinfo {year} {1985})}\BibitemShut {NoStop}%
\bibitem [{\citenamefont {Sivardi{\`{e}}re}(1988)}]{Sivardiere1988}%
  \BibitemOpen
  \bibfield  {author} {\bibinfo {author} {\bibfnamefont {J.}~\bibnamefont {Sivardi{\`{e}}re}},\ }\href {\doibase 10.1107/S0108767388005549} {\bibfield  {journal} {\bibinfo  {journal} {Acta Crystallographica}\ }\textbf {\bibinfo {volume} {44}},\ \bibinfo {pages} {735} (\bibinfo {year} {1988})}\BibitemShut {NoStop}%
\bibitem [{\citenamefont {Kotzev}\ and\ \citenamefont {Alexandrova}(1988)}]{Kotzev1988}%
  \BibitemOpen
  \bibfield  {author} {\bibinfo {author} {\bibfnamefont {J.~N.}\ \bibnamefont {Kotzev}}\ and\ \bibinfo {author} {\bibfnamefont {D.~A.}\ \bibnamefont {Alexandrova}},\ }\href {\doibase 10.1107/S0108767388008335} {\bibfield  {journal} {\bibinfo  {journal} {Acta Crystallographica Section A}\ }\textbf {\bibinfo {volume} {44}},\ \bibinfo {pages} {1082} (\bibinfo {year} {1988})}\BibitemShut {NoStop}%
\bibitem [{\citenamefont {Litvin}\ \emph {et~al.}(1982)\citenamefont {Litvin}, \citenamefont {Kotzev},\ and\ \citenamefont {Birman}}]{Litvin1982}%
  \BibitemOpen
  \bibfield  {author} {\bibinfo {author} {\bibfnamefont {D.~B.}\ \bibnamefont {Litvin}}, \bibinfo {author} {\bibfnamefont {J.~N.}\ \bibnamefont {Kotzev}}, \ and\ \bibinfo {author} {\bibfnamefont {J.~L.}\ \bibnamefont {Birman}},\ }\href {\doibase 10.1103/PhysRevB.26.6947} {\bibfield  {journal} {\bibinfo  {journal} {Physical Review B}\ }\textbf {\bibinfo {volume} {26}},\ \bibinfo {pages} {6947} (\bibinfo {year} {1982})}\BibitemShut {NoStop}%
\bibitem [{\citenamefont {Jir{\'{a}}k}(1992)}]{Jirak1992}%
  \BibitemOpen
  \bibfield  {author} {\bibinfo {author} {\bibfnamefont {Z.}~\bibnamefont {Jir{\'{a}}k}},\ }\href {\doibase 10.1103/PhysRevB.46.8725} {\bibfield  {journal} {\bibinfo  {journal} {Physical Review B}\ }\textbf {\bibinfo {volume} {46}},\ \bibinfo {pages} {8725} (\bibinfo {year} {1992})}\BibitemShut {NoStop}%
\bibitem [{\citenamefont {Bertaut}(1968)}]{Bertaut1968}%
  \BibitemOpen
  \bibfield  {author} {\bibinfo {author} {\bibfnamefont {E.~F.}\ \bibnamefont {Bertaut}},\ }\href {\doibase 10.1107/S0567739468000306} {\bibfield  {journal} {\bibinfo  {journal} {Acta Crystallographica Section A}\ }\textbf {\bibinfo {volume} {24}},\ \bibinfo {pages} {217} (\bibinfo {year} {1968})}\BibitemShut {NoStop}%
\bibitem [{\citenamefont {Izyumov}(1980)}]{Izyumov1980}%
  \BibitemOpen
  \bibfield  {author} {\bibinfo {author} {\bibfnamefont {Y.~A.}\ \bibnamefont {Izyumov}},\ }\href {\doibase 10.1016/0304-8853(80)91147-6} {\bibfield  {journal} {\bibinfo  {journal} {Journal of Magnetism and Magnetic Materials}\ }\textbf {\bibinfo {volume} {15-18}},\ \bibinfo {pages} {497} (\bibinfo {year} {1980})}\BibitemShut {NoStop}%
\bibitem [{\citenamefont {Litvin}(1977)}]{Litvin1977}%
  \BibitemOpen
  \bibfield  {author} {\bibinfo {author} {\bibfnamefont {D.~B.}\ \bibnamefont {Litvin}},\ }\href {\doibase 10.1107/S0567739477000709} {\bibfield  {journal} {\bibinfo  {journal} {Acta Crystallographica Section A}\ }\textbf {\bibinfo {volume} {33}},\ \bibinfo {pages} {279} (\bibinfo {year} {1977})}\BibitemShut {NoStop}%
\bibitem [{\citenamefont {Eriksson}\ \emph {et~al.}(2004{\natexlab{a}})\citenamefont {Eriksson}, \citenamefont {Bergqvist}, \citenamefont {Nordblad}, \citenamefont {Eriksson},\ and\ \citenamefont {Andersson}}]{Eriksson2004a}%
  \BibitemOpen
  \bibfield  {author} {\bibinfo {author} {\bibfnamefont {T.}~\bibnamefont {Eriksson}}, \bibinfo {author} {\bibfnamefont {L.}~\bibnamefont {Bergqvist}}, \bibinfo {author} {\bibfnamefont {P.}~\bibnamefont {Nordblad}}, \bibinfo {author} {\bibfnamefont {O.}~\bibnamefont {Eriksson}}, \ and\ \bibinfo {author} {\bibfnamefont {Y.}~\bibnamefont {Andersson}},\ }\href {\doibase 10.1016/J.JSSC.2004.07.001} {\bibfield  {journal} {\bibinfo  {journal} {Journal of Solid State Chemistry}\ }\textbf {\bibinfo {volume} {177}},\ \bibinfo {pages} {4058} (\bibinfo {year} {2004}{\natexlab{a}})}\BibitemShut {NoStop}%
\bibitem [{\citenamefont {Eriksson}\ \emph {et~al.}(2004{\natexlab{b}})\citenamefont {Eriksson}, \citenamefont {Liz{\'{a}}rraga}, \citenamefont {Felton}, \citenamefont {Bergqvist}, \citenamefont {Andersson}, \citenamefont {Nordblad},\ and\ \citenamefont {Eriksson}}]{Eriksson2004}%
  \BibitemOpen
  \bibfield  {author} {\bibinfo {author} {\bibfnamefont {T.}~\bibnamefont {Eriksson}}, \bibinfo {author} {\bibfnamefont {R.}~\bibnamefont {Liz{\'{a}}rraga}}, \bibinfo {author} {\bibfnamefont {S.}~\bibnamefont {Felton}}, \bibinfo {author} {\bibfnamefont {L.}~\bibnamefont {Bergqvist}}, \bibinfo {author} {\bibfnamefont {Y.}~\bibnamefont {Andersson}}, \bibinfo {author} {\bibfnamefont {P.}~\bibnamefont {Nordblad}}, \ and\ \bibinfo {author} {\bibfnamefont {O.}~\bibnamefont {Eriksson}},\ }\href {\doibase 10.1103/PhysRevB.69.054422} {\bibfield  {journal} {\bibinfo  {journal} {Physical Review B}\ }\textbf {\bibinfo {volume} {69}},\ \bibinfo {pages} {054422} (\bibinfo {year} {2004}{\natexlab{b}})}\BibitemShut {NoStop}%
\bibitem [{\citenamefont {Hu}\ \emph {et~al.}(2024)\citenamefont {Hu}, \citenamefont {Janson}, \citenamefont {Felser}, \citenamefont {McClarty}, \citenamefont {van~den Brink},\ and\ \citenamefont {Vergniory}}]{Hu2024}%
  \BibitemOpen
  \bibfield  {author} {\bibinfo {author} {\bibfnamefont {M.}~\bibnamefont {Hu}}, \bibinfo {author} {\bibfnamefont {O.}~\bibnamefont {Janson}}, \bibinfo {author} {\bibfnamefont {C.}~\bibnamefont {Felser}}, \bibinfo {author} {\bibfnamefont {P.}~\bibnamefont {McClarty}}, \bibinfo {author} {\bibfnamefont {J.}~\bibnamefont {van~den Brink}}, \ and\ \bibinfo {author} {\bibfnamefont {M.~G.}\ \bibnamefont {Vergniory}},\ }\href {https://arxiv.org/abs/2410.17993v2} {\  (\bibinfo {year} {2024})},\ \Eprint {http://arxiv.org/abs/2410.17993} {arXiv:2410.17993} \BibitemShut {NoStop}%
\bibitem [{\citenamefont {Kimber}\ and\ \citenamefont {Attfield}(2007)}]{Kimber2007}%
  \BibitemOpen
  \bibfield  {author} {\bibinfo {author} {\bibfnamefont {S.~A.}\ \bibnamefont {Kimber}}\ and\ \bibinfo {author} {\bibfnamefont {J.~P.}\ \bibnamefont {Attfield}},\ }\href {\doibase 10.1039/B704361A} {\bibfield  {journal} {\bibinfo  {journal} {Journal of Materials Chemistry}\ }\textbf {\bibinfo {volume} {17}},\ \bibinfo {pages} {4885} (\bibinfo {year} {2007})}\BibitemShut {NoStop}%
\bibitem [{\citenamefont {Kakarla}\ \emph {et~al.}(2019)\citenamefont {Kakarla}, \citenamefont {Wu}, \citenamefont {Hsieh}, \citenamefont {Sun}, \citenamefont {Dai}, \citenamefont {Lin}, \citenamefont {Her}, \citenamefont {Matsuda}, \citenamefont {Deng}, \citenamefont {Gooch}, \citenamefont {Chu},\ and\ \citenamefont {Yang}}]{Kakarla2019}%
  \BibitemOpen
  \bibfield  {author} {\bibinfo {author} {\bibfnamefont {D.~C.}\ \bibnamefont {Kakarla}}, \bibinfo {author} {\bibfnamefont {H.~C.}\ \bibnamefont {Wu}}, \bibinfo {author} {\bibfnamefont {D.~J.}\ \bibnamefont {Hsieh}}, \bibinfo {author} {\bibfnamefont {P.~J.}\ \bibnamefont {Sun}}, \bibinfo {author} {\bibfnamefont {G.~J.}\ \bibnamefont {Dai}}, \bibinfo {author} {\bibfnamefont {J.~Y.}\ \bibnamefont {Lin}}, \bibinfo {author} {\bibfnamefont {J.~L.}\ \bibnamefont {Her}}, \bibinfo {author} {\bibfnamefont {Y.~H.}\ \bibnamefont {Matsuda}}, \bibinfo {author} {\bibfnamefont {L.~Z.}\ \bibnamefont {Deng}}, \bibinfo {author} {\bibfnamefont {M.}~\bibnamefont {Gooch}}, \bibinfo {author} {\bibfnamefont {C.~W.}\ \bibnamefont {Chu}}, \ and\ \bibinfo {author} {\bibfnamefont {H.~D.}\ \bibnamefont {Yang}},\ }\href {\doibase 10.1103/PHYSREVB.99.195129/SUPPLEMENTARY_MATERIAL-B13484.PDF} {\bibfield  {journal} {\bibinfo  {journal} {Physical Review B}\ }\textbf {\bibinfo {volume} {99}},\ \bibinfo {pages} {195129} (\bibinfo {year}
  {2019})}\BibitemShut {NoStop}%
\bibitem [{\citenamefont {Cheong}\ and\ \citenamefont {Huang}(2025)}]{Cheong2025}%
  \BibitemOpen
  \bibfield  {author} {\bibinfo {author} {\bibfnamefont {S.-W.}\ \bibnamefont {Cheong}}\ and\ \bibinfo {author} {\bibfnamefont {F.-T.}\ \bibnamefont {Huang}},\ }\href {https://arxiv.org/abs/2503.16277v1} {\  (\bibinfo {year} {2025})},\ \Eprint {http://arxiv.org/abs/2503.16277} {arXiv:2503.16277} \BibitemShut {NoStop}%
\bibitem [{\citenamefont {Bertaut}\ \emph {et~al.}(1968)\citenamefont {Bertaut}, \citenamefont {Fruchart}, \citenamefont {Bouchaud},\ and\ \citenamefont {Fruchart}}]{Bertaut1968b}%
  \BibitemOpen
  \bibfield  {author} {\bibinfo {author} {\bibfnamefont {E.~F.}\ \bibnamefont {Bertaut}}, \bibinfo {author} {\bibfnamefont {D.}~\bibnamefont {Fruchart}}, \bibinfo {author} {\bibfnamefont {J.~P.}\ \bibnamefont {Bouchaud}}, \ and\ \bibinfo {author} {\bibfnamefont {R.}~\bibnamefont {Fruchart}},\ }\href {\doibase 10.1016/0038-1098(68)90098-7} {\bibfield  {journal} {\bibinfo  {journal} {Solid State Communications}\ }\textbf {\bibinfo {volume} {6}},\ \bibinfo {pages} {251} (\bibinfo {year} {1968})}\BibitemShut {NoStop}%
\bibitem [{\citenamefont {Shi}\ \emph {et~al.}(2016)\citenamefont {Shi}, \citenamefont {Sun}, \citenamefont {Yan}, \citenamefont {Deng}, \citenamefont {Wang}, \citenamefont {Wu}, \citenamefont {Hu}, \citenamefont {Lu}, \citenamefont {Malik}, \citenamefont {Huang},\ and\ \citenamefont {Wang}}]{Shi2016}%
  \BibitemOpen
  \bibfield  {author} {\bibinfo {author} {\bibfnamefont {K.}~\bibnamefont {Shi}}, \bibinfo {author} {\bibfnamefont {Y.}~\bibnamefont {Sun}}, \bibinfo {author} {\bibfnamefont {J.}~\bibnamefont {Yan}}, \bibinfo {author} {\bibfnamefont {S.}~\bibnamefont {Deng}}, \bibinfo {author} {\bibfnamefont {L.}~\bibnamefont {Wang}}, \bibinfo {author} {\bibfnamefont {H.}~\bibnamefont {Wu}}, \bibinfo {author} {\bibfnamefont {P.}~\bibnamefont {Hu}}, \bibinfo {author} {\bibfnamefont {H.}~\bibnamefont {Lu}}, \bibinfo {author} {\bibfnamefont {M.~I.}\ \bibnamefont {Malik}}, \bibinfo {author} {\bibfnamefont {Q.}~\bibnamefont {Huang}}, \ and\ \bibinfo {author} {\bibfnamefont {C.}~\bibnamefont {Wang}},\ }\href {\doibase 10.1002/ADMA.201600310} {\bibfield  {journal} {\bibinfo  {journal} {Advanced materials (Deerfield Beach, Fla.)}\ }\textbf {\bibinfo {volume} {28}},\ \bibinfo {pages} {3761} (\bibinfo {year} {2016})}\BibitemShut {NoStop}%
\bibitem [{\citenamefont {Nan}\ \emph {et~al.}(2020)\citenamefont {Nan}, \citenamefont {Quintela}, \citenamefont {Irwin}, \citenamefont {Gurung}, \citenamefont {Shao}, \citenamefont {Gibbons}, \citenamefont {Campbell}, \citenamefont {Song}, \citenamefont {Choi}, \citenamefont {Guo}, \citenamefont {Johnson}, \citenamefont {Manuel}, \citenamefont {Chopdekar}, \citenamefont {Hallsteinsen}, \citenamefont {Tybell}, \citenamefont {Ryan}, \citenamefont {Kim}, \citenamefont {Choi}, \citenamefont {Radaelli}, \citenamefont {Ralph}, \citenamefont {Tsymbal}, \citenamefont {Rzchowski},\ and\ \citenamefont {Eom}}]{Nan2020}%
  \BibitemOpen
  \bibfield  {author} {\bibinfo {author} {\bibfnamefont {T.}~\bibnamefont {Nan}}, \bibinfo {author} {\bibfnamefont {C.~X.}\ \bibnamefont {Quintela}}, \bibinfo {author} {\bibfnamefont {J.}~\bibnamefont {Irwin}}, \bibinfo {author} {\bibfnamefont {G.}~\bibnamefont {Gurung}}, \bibinfo {author} {\bibfnamefont {D.~F.}\ \bibnamefont {Shao}}, \bibinfo {author} {\bibfnamefont {J.}~\bibnamefont {Gibbons}}, \bibinfo {author} {\bibfnamefont {N.}~\bibnamefont {Campbell}}, \bibinfo {author} {\bibfnamefont {K.}~\bibnamefont {Song}}, \bibinfo {author} {\bibfnamefont {S.~Y.}\ \bibnamefont {Choi}}, \bibinfo {author} {\bibfnamefont {L.}~\bibnamefont {Guo}}, \bibinfo {author} {\bibfnamefont {R.~D.}\ \bibnamefont {Johnson}}, \bibinfo {author} {\bibfnamefont {P.}~\bibnamefont {Manuel}}, \bibinfo {author} {\bibfnamefont {R.~V.}\ \bibnamefont {Chopdekar}}, \bibinfo {author} {\bibfnamefont {I.}~\bibnamefont {Hallsteinsen}}, \bibinfo {author} {\bibfnamefont {T.}~\bibnamefont {Tybell}}, \bibinfo {author} {\bibfnamefont {P.~J.}\
  \bibnamefont {Ryan}}, \bibinfo {author} {\bibfnamefont {J.~W.}\ \bibnamefont {Kim}}, \bibinfo {author} {\bibfnamefont {Y.}~\bibnamefont {Choi}}, \bibinfo {author} {\bibfnamefont {P.~G.}\ \bibnamefont {Radaelli}}, \bibinfo {author} {\bibfnamefont {D.~C.}\ \bibnamefont {Ralph}}, \bibinfo {author} {\bibfnamefont {E.~Y.}\ \bibnamefont {Tsymbal}}, \bibinfo {author} {\bibfnamefont {M.~S.}\ \bibnamefont {Rzchowski}}, \ and\ \bibinfo {author} {\bibfnamefont {C.~B.}\ \bibnamefont {Eom}},\ }\href {\doibase 10.1038/s41467-020-17999-4} {\bibfield  {journal} {\bibinfo  {journal} {Nature Communications 2020 11:1}\ }\textbf {\bibinfo {volume} {11}},\ \bibinfo {pages} {1} (\bibinfo {year} {2020})},\ \Eprint {http://arxiv.org/abs/1912.12586} {arXiv:1912.12586} \BibitemShut {NoStop}%
\bibitem [{\citenamefont {Singh}\ \emph {et~al.}(2021)\citenamefont {Singh}, \citenamefont {Samathrakis}, \citenamefont {Fortunato}, \citenamefont {Zemen}, \citenamefont {Shen}, \citenamefont {Gutfleisch},\ and\ \citenamefont {Zhang}}]{Singh2021}%
  \BibitemOpen
  \bibfield  {author} {\bibinfo {author} {\bibfnamefont {H.~K.}\ \bibnamefont {Singh}}, \bibinfo {author} {\bibfnamefont {I.}~\bibnamefont {Samathrakis}}, \bibinfo {author} {\bibfnamefont {N.~M.}\ \bibnamefont {Fortunato}}, \bibinfo {author} {\bibfnamefont {J.}~\bibnamefont {Zemen}}, \bibinfo {author} {\bibfnamefont {C.}~\bibnamefont {Shen}}, \bibinfo {author} {\bibfnamefont {O.}~\bibnamefont {Gutfleisch}}, \ and\ \bibinfo {author} {\bibfnamefont {H.}~\bibnamefont {Zhang}},\ }\href {\doibase 10.1038/s41524-021-00566-w} {\bibfield  {journal} {\bibinfo  {journal} {npj Computational Materials 2021 7:1}\ }\textbf {\bibinfo {volume} {7}},\ \bibinfo {pages} {1} (\bibinfo {year} {2021})},\ \Eprint {http://arxiv.org/abs/2009.06440} {arXiv:2009.06440} \BibitemShut {NoStop}%
\bibitem [{Note3()}]{Note3}%
  \BibitemOpen
  \bibinfo {note} {Note that the $\Gamma ^{4g}$ magnetic structure is the same as for the compound Mn$_3$Ir reported in ref. \protect \rev@citealpnum {Zelezny2017}. The schematic spin texture reported in their Fig. 3 is very similar to the one we calculate here.}\BibitemShut {Stop}%
\bibitem [{\citenamefont {Giannozzi}\ \emph {et~al.}(2017)\citenamefont {Giannozzi}, \citenamefont {Andreussi}, \citenamefont {Brumme}, \citenamefont {Bunau}, \citenamefont {{Buongiorno Nardelli}}, \citenamefont {Calandra}, \citenamefont {Car}, \citenamefont {Cavazzoni}, \citenamefont {Ceresoli}, \citenamefont {Cococcioni}, \citenamefont {Colonna}, \citenamefont {Carnimeo}, \citenamefont {{Dal Corso}}, \citenamefont {{De Gironcoli}}, \citenamefont {Delugas}, \citenamefont {Distasio}, \citenamefont {Ferretti}, \citenamefont {Floris}, \citenamefont {Fratesi}, \citenamefont {Fugallo}, \citenamefont {Gebauer}, \citenamefont {Gerstmann}, \citenamefont {Giustino}, \citenamefont {Gorni}, \citenamefont {Jia}, \citenamefont {Kawamura}, \citenamefont {Ko}, \citenamefont {Kokalj}, \citenamefont {K{\"{u}}c{\"{u}}kbenli}, \citenamefont {Lazzeri}, \citenamefont {Marsili}, \citenamefont {Marzari}, \citenamefont {Mauri}, \citenamefont {Nguyen}, \citenamefont {Nguyen}, \citenamefont {Otero-De-La-Roza}, \citenamefont
  {Paulatto}, \citenamefont {Ponc{\'{e}}}, \citenamefont {Rocca}, \citenamefont {Sabatini}, \citenamefont {Santra}, \citenamefont {Schlipf}, \citenamefont {Seitsonen}, \citenamefont {Smogunov}, \citenamefont {Timrov}, \citenamefont {Thonhauser}, \citenamefont {Umari}, \citenamefont {Vast}, \citenamefont {Wu},\ and\ \citenamefont {Baroni}}]{QE-2017}%
  \BibitemOpen
  \bibfield  {author} {\bibinfo {author} {\bibfnamefont {P.}~\bibnamefont {Giannozzi}}, \bibinfo {author} {\bibfnamefont {O.}~\bibnamefont {Andreussi}}, \bibinfo {author} {\bibfnamefont {T.}~\bibnamefont {Brumme}}, \bibinfo {author} {\bibfnamefont {O.}~\bibnamefont {Bunau}}, \bibinfo {author} {\bibfnamefont {M.}~\bibnamefont {{Buongiorno Nardelli}}}, \bibinfo {author} {\bibfnamefont {M.}~\bibnamefont {Calandra}}, \bibinfo {author} {\bibfnamefont {R.}~\bibnamefont {Car}}, \bibinfo {author} {\bibfnamefont {C.}~\bibnamefont {Cavazzoni}}, \bibinfo {author} {\bibfnamefont {D.}~\bibnamefont {Ceresoli}}, \bibinfo {author} {\bibfnamefont {M.}~\bibnamefont {Cococcioni}}, \bibinfo {author} {\bibfnamefont {N.}~\bibnamefont {Colonna}}, \bibinfo {author} {\bibfnamefont {I.}~\bibnamefont {Carnimeo}}, \bibinfo {author} {\bibfnamefont {A.}~\bibnamefont {{Dal Corso}}}, \bibinfo {author} {\bibfnamefont {S.}~\bibnamefont {{De Gironcoli}}}, \bibinfo {author} {\bibfnamefont {P.}~\bibnamefont {Delugas}}, \bibinfo {author}
  {\bibfnamefont {R.~A.}\ \bibnamefont {Distasio}}, \bibinfo {author} {\bibfnamefont {A.}~\bibnamefont {Ferretti}}, \bibinfo {author} {\bibfnamefont {A.}~\bibnamefont {Floris}}, \bibinfo {author} {\bibfnamefont {G.}~\bibnamefont {Fratesi}}, \bibinfo {author} {\bibfnamefont {G.}~\bibnamefont {Fugallo}}, \bibinfo {author} {\bibfnamefont {R.}~\bibnamefont {Gebauer}}, \bibinfo {author} {\bibfnamefont {U.}~\bibnamefont {Gerstmann}}, \bibinfo {author} {\bibfnamefont {F.}~\bibnamefont {Giustino}}, \bibinfo {author} {\bibfnamefont {T.}~\bibnamefont {Gorni}}, \bibinfo {author} {\bibfnamefont {J.}~\bibnamefont {Jia}}, \bibinfo {author} {\bibfnamefont {M.}~\bibnamefont {Kawamura}}, \bibinfo {author} {\bibfnamefont {H.~Y.}\ \bibnamefont {Ko}}, \bibinfo {author} {\bibfnamefont {A.}~\bibnamefont {Kokalj}}, \bibinfo {author} {\bibfnamefont {E.}~\bibnamefont {K{\"{u}}c{\"{u}}kbenli}}, \bibinfo {author} {\bibfnamefont {M.}~\bibnamefont {Lazzeri}}, \bibinfo {author} {\bibfnamefont {M.}~\bibnamefont {Marsili}}, \bibinfo
  {author} {\bibfnamefont {N.}~\bibnamefont {Marzari}}, \bibinfo {author} {\bibfnamefont {F.}~\bibnamefont {Mauri}}, \bibinfo {author} {\bibfnamefont {N.~L.}\ \bibnamefont {Nguyen}}, \bibinfo {author} {\bibfnamefont {H.~V.}\ \bibnamefont {Nguyen}}, \bibinfo {author} {\bibfnamefont {A.}~\bibnamefont {Otero-De-La-Roza}}, \bibinfo {author} {\bibfnamefont {L.}~\bibnamefont {Paulatto}}, \bibinfo {author} {\bibfnamefont {S.}~\bibnamefont {Ponc{\'{e}}}}, \bibinfo {author} {\bibfnamefont {D.}~\bibnamefont {Rocca}}, \bibinfo {author} {\bibfnamefont {R.}~\bibnamefont {Sabatini}}, \bibinfo {author} {\bibfnamefont {B.}~\bibnamefont {Santra}}, \bibinfo {author} {\bibfnamefont {M.}~\bibnamefont {Schlipf}}, \bibinfo {author} {\bibfnamefont {A.~P.}\ \bibnamefont {Seitsonen}}, \bibinfo {author} {\bibfnamefont {A.}~\bibnamefont {Smogunov}}, \bibinfo {author} {\bibfnamefont {I.}~\bibnamefont {Timrov}}, \bibinfo {author} {\bibfnamefont {T.}~\bibnamefont {Thonhauser}}, \bibinfo {author} {\bibfnamefont {P.}~\bibnamefont {Umari}},
  \bibinfo {author} {\bibfnamefont {N.}~\bibnamefont {Vast}}, \bibinfo {author} {\bibfnamefont {X.}~\bibnamefont {Wu}}, \ and\ \bibinfo {author} {\bibfnamefont {S.}~\bibnamefont {Baroni}},\ }\href {\doibase 10.1088/1361-648X/AA8F79} {\bibfield  {journal} {\bibinfo  {journal} {Journal of Physics: Condensed Matter}\ }\textbf {\bibinfo {volume} {29}},\ \bibinfo {pages} {465901} (\bibinfo {year} {2017})},\ \Eprint {http://arxiv.org/abs/1709.10010} {arXiv:1709.10010} \BibitemShut {NoStop}%
\bibitem [{\citenamefont {Perdew}\ \emph {et~al.}(1996)\citenamefont {Perdew}, \citenamefont {Burke},\ and\ \citenamefont {Ernzerhof}}]{GGA}%
  \BibitemOpen
  \bibfield  {author} {\bibinfo {author} {\bibfnamefont {J.~P.}\ \bibnamefont {Perdew}}, \bibinfo {author} {\bibfnamefont {K.}~\bibnamefont {Burke}}, \ and\ \bibinfo {author} {\bibfnamefont {M.}~\bibnamefont {Ernzerhof}},\ }\href {\doibase 10.1103/PhysRevLett.77.3865} {\bibfield  {journal} {\bibinfo  {journal} {Phys. Rev. Lett.}\ }\textbf {\bibinfo {volume} {77}},\ \bibinfo {pages} {3865} (\bibinfo {year} {1996})}\BibitemShut {NoStop}%
\bibitem [{sup()}]{supp}%
  \BibitemOpen
  \href@noop {} {}\bibinfo {note} {See Supplemental Material.}\BibitemShut {Stop}%
\bibitem [{Note4()}]{Note4}%
  \BibitemOpen
  \bibinfo {note} {Note that the net magnetization in the absence of SOC is \protect \emph {identically zero} for collinear altermagnets and also for non-collinear magnets where each colour has a corresponding anti-colour, since the magnetisation of anti-coloured sublattices cancels.}\BibitemShut {Stop}%
\bibitem [{Note5()}]{Note5}%
  \BibitemOpen
  \bibinfo {note} {Very briefly, magnets with $\protect \mathbf {k}/-\protect \mathbf {k}$ \protect \emph {anti-symmetric} splitting are characterised by bond-type multipolar ordering,\cite {Hayami2020b} and the staggered magnetisation (in the collinear case) or the one-colour axial unit vectors (non-collinear case) need to be replaced in the tensorial treatment by the cross product $\protect \bm {S}_1 \times \protect \bm {S}_2$, where sites 1 and 2 are those connected by the specific bond. This approach will be fully described in a future paper.}\BibitemShut {Stop}%
\end{thebibliography}%

\end{document}